\documentclass{aastex631}

\begin{document}

\title{

Investigating the impact of atomic data uncertainties on measured physical parameters of the Perseus galaxy cluster }

\correspondingauthor{Priyanka Chakraborty}
\email{priyanka.chakraborty@cfa.harvard.edu}

\author[0000-0002-4469-2518]{Priyanka Chakraborty}
\affiliation{Center for Astrophysics $|$ Harvard \& Smithsonian  \\
60 Garden St. \\
Cambridge, MA 02138, USA}

\author[0000-0002-6449-9299]{Rachel Hemmer}
\affiliation{Brown University \\
182 Hope Street, Box 1843 \\
Providence, RI, 02912, USA}


\author[0000-0003-3462-8886]{Adam R. Foster}
\affiliation{Center for Astrophysics $|$ Harvard \& Smithsonian  \\
60 Garden St. \\
Cambridge, MA 02138, USA}

\author[0000-0002-7868-1622]{John Raymond}
\affiliation{Center for Astrophysics $|$ Harvard \& Smithsonian  \\
60 Garden St. \\
Cambridge, MA 02138, USA}

\author[0000-0002-5222-1337]{Arnab Sarkar}
\affiliation{Department of Physics, Kavli Institute for Astrophysics and Space Research, \\Massachusetts Institute of Technology, Cambridge, MA 02139, USA\\
}

\author[0000-0003-4284-4167]{Randall Smith}
\affiliation{Center for Astrophysics $|$ Harvard \& Smithsonian  \\
60 Garden St. \\
Cambridge, MA 02138, USA}

\author[0000-0002-8704-4473]{Nancy Brickhouse}
\affiliation{Center for Astrophysics $|$ Harvard \& Smithsonian  \\
60 Garden St. \\
Cambridge, MA 02138, USA}





\begin{abstract}

Accurate atomic data and plasma models are essential for interpreting the upcoming high-quality spectra from missions like \textit{XRISM} and \textit{Athena}. Estimating physical quantities, like temperature, abundance, turbulence, and resonance scattering factor, is highly dependent on the underlying atomic data. We use the AtomDB tool \texttt{variableapec} to estimate the impact of atomic data uncertainties in Einstein A coefficients, collisional rate coefficient, ionization and recombination rates in H-, He- and Li-like iron in modeling the spectrum of Perseus observed by \textit{Hitomi}. The best-fit temperature, abundance, resonance scattering factor, and turbulence including atomic data uncertainties, varied approximately 17\%,  35\%,  30\%, and  3\%, respectively, from the best-fit temperature, abundance, resonance scattering factor, and turbulence estimated without atomic data uncertainties. This indicates that, approximately 32\%, 35\%, and  25\% of the best-fit temperatures, abundances, and resonance scattering factors, including uncertainties lie outside the 3$\sigma$ error regions of their corresponding best-fit values computed with zero atomic data error.
Expanding the energy range to 1.8-20.0 keV shows less variability,  with 26\% of the abundances and 22\% of the resonance scattering factors lying outside the 3$\sigma$ error of the best-fit values.

We also studied correlations between physical parameters and atomic rate uncertainties to identify key atomic quantities requiring precise lab measurements. We report negative correlations between best-fit temperature and z (1s.2s $^{3}\rm S_{1}\rightarrow 1s^{2}$) collisional rate coefficient, abundance and y (1s.2p $^{3}\rm P_{1}\rightarrow 1s^{2}$) collisional rate coefficient, abundance and z collisional rate coefficient, and positive correlation between resonance scattering factor and w (1s.2p $^{1}\rm P_{1}\rightarrow 1s^{2}$) collisional rate coefficient.


\end{abstract}

\keywords{}


\section{Introduction} \label{sec:intro}
The upcoming microcalorimeter missions \textit{XRISM} and \textit{Athena}
will perform high-resolution spectroscopy of a broad range of astrophysical sources, providing unprecedented insight into the nature of the high-energy universe. 
Interpretation of such high-quality X-ray observation is heavily reliant on the accuracy of the underlying fundamental data used in spectral synthesis codes like APEC \citep{2001ApJ...556L..91S, 2012ApJ...756..128F}. For example, the uncertainty in atomic data can bias the measurement of various parameters of galaxy clusters and groups, such as their temperature, metallicity, and density, leading to incorrect conclusions about the properties of the Intra-Cluster Medium (ICM)
and enrichment history of the cluster as a whole \citep{2018SSRv..214..129M, 2018MNRAS.478L.116M, 2020AN....341..203M,2022MNRAS.516.3068S,2023arXiv231004499M}. Chemical abundances in supernova remnants (SNRs) are
sensitive to the atomic data uncertainties with crucial implications for the interpretation of the nucleosynthesis processes occurring in supernova explosions 
\citep{2017ApJ...845..176B, 2018SSRv..214...67N}.  Uncertainties in atomic data, particularly for iron-group elements, can lead to systematic errors in the inferred stellar parameters, such as effective temperature, surface gravity, and metallicity, important for understanding stellar properties and evolution \citep{2018ApJ...866..146Y}. The precision  and reliability of  scattering measurements, such as resonance scattering (RS) and electron scattering escape \citep{2020ApJ...901...68C, 2020RNAAS...4..184C, 2021ApJ...912...26C} are also sensitive to the accuracy of the atomic data.

Various line-ratio diagnostics using spectral synthesis softwares combined with future microcalorimeter measurements will enable us to map the physical properties of X-ray emitting plasma, including temperature structure, abundance, and velocity dispersion \citep[e.g.,][]{2020ApJ...901...69C, 2022ApJ...935...70C, 2022arXiv221109827K, 2023arXiv230701277T, 2023arXiv230701259S, 2023arXiv231002225Z, 2023MNRAS.522.3665N}. 
Over the past few years, advances in spectroscopic instruments and theoretical models have led to significant improvements in the accuracy of  the underlying atomic data and models used in such spectral synthesis, frequently achieving confidence levels of up to 95\% \citep{2014Atoms...2...86K}.  Yet 
in the absence of decisive experimental measurements, it is not clear that one of
the data sets used in the atomic models is more correct than the other. 
Recently, 
the high-resolution \textit{Hitomi} spectra of the Perseus cluster revealed
that the ambiguity in the interpretation of physical parameters
from the data often stems from the collisional-radiative models used in the analysis \citep{model_constraints}. Statistically significant discrepancies were found  in the plasma
parameters in the analysis carried out with APEC and SPEX.

For collisional astrophysical plasma, spectral line intensities are  sensitive to the accuracy in  Einstein A coefficients  (A values),  collisional rate coefficients representing the Maxwellian average of the collision cross-section (q values),
collisional ionization (CI) rates, dielectronic recombination (DR) rates, and radiative recombination (RR) rates  of each ion/their adjacent ions \citep{2019MNRAS.484.4754D}.
In this paper, we explore the impact of uncertainties in the above atomic datasets in constraining physical parameters/processes like temperature, elemental abundance,   turbulence, and resonance scattering (RS)\footnote{Resonance scattering distorts X-ray spectral lines by redirecting photons, resulting in intensity reductions in certain areas (like galaxy centers) and enhancements in others (such as outskirts). This occurs when optically thick line photons are absorbed and promptly re-emitted by ions with similar resonant transition energies, causing a change in the line emission distribution.} using the \textit{Hitomi} spectrum of Perseus core.  We used the latest version of AtomDB (3.0.9)  to model the hot collisional plasma of Perseus and produce line emissivities (and ratios) for selected He- and Li-like transitions for a range of uncertainties obtained from various atomic databases like SPEX \citep{1996uxsa.conf..411K},   AtomDB \citep{2001ApJ...556L..91S, 2012ApJ...756..128F},  CHIANTI \citep{1997A&AS..125..149D,2015A&A...582A..56D},  FAC \citep{doi:10.1139/p07-197},  Open-ADAS \footnote{ADF04, OPEN-ADAS database, Website: \url{https://open.adas.ac.uk/}}, and NIST \citep{2018APS..DMPM01004K}. All our calculations were done using the tool \texttt{variableapec} \citep{2021ApJ...908....3H} based on AtomDB, which now offers the flexibility of varying atomic rates for individual transitions. We answer the following questions:
a) Which line emissivities/line ratios are sensitive to the uncertainty in which fundamental atomic data, if at all, and to what extent? b) Are the physical parameters derived from fitting the observed spectrum of the Perseus cluster sensitive to the uncertainty in atomic data, and to what degree? c) Are there any correlations between the measured physical parameters with individual atomic rates? Answering the former two will identify
 the extent to which the \textit{XRISM}'s scientific goals might be limited by the accuracy of the fundamental atomic data used in spectral modeling. Answering the latter will 
motivate precise laboratory measurements of certain atomic quantities tightly correlated with the measured physical parameters. In this paper, we concentrate our spectral fitting on the 6.0-6.8 keV range, incorporating both Fe XXV and Fe XXIV lines. Extending the fit to the full 1.8-20.0 keV spectrum requires understanding uncertainties in transitions like Si, S, Ar, and Ca ions across databases. We plan to address this comprehensively in a future paper.

Several techniques exist to estimate uncertainties, such as experimental error estimation, for example, errors associated with electron beam ion trap (EBIT) experiments,  systematic errors (e.g., instrument calibration issues)
 random errors (e.g., measurement fluctuations), comparison with observed spectra like that of Capella, and theoretical error estimation through statistical techniques like Bayesian analysis or 
incorporating advancements in radiative transfer and computational methods. However, it is not straightforward to decide which technique is the best.
As a result, there have been very few atomic benchmarking studies quantifying the effect of data uncertainty in atomic spectroscopy \citep{2017HEDP...23...38P}.  In this paper, for the first time, we present a  technique/tool for quantifying such effects and list the uncertainties in the calculated physical parameters in the context of high-resolution X-ray spectroscopy.

\begin{table*}
\centering{    
\caption{ List of relevant Fe$^{24+}$ and Fe$^{23+}$ lines around the Fe K$\alpha$ complex sensitive to the changes in Einstein A coefficients  (s$^{-1}$) and  collisional rate coefficients  (cm$^{3}$s$^{-1}$), along with their percentage uncertainties. Maximum error percentages in the Fe$^{24+}$ lines for Einstein A coefficients  (A values) and  collisional rate coefficients (q values) have been calculated comparing SPEX, AtomDB, CHIANTI,  FAC, and Open-ADAS databases and taken from \citet{model_constraints}. The Fe$^{23+}$ Einstein A coefficients  are taken from the NIST database, reported within 25\% accuracy. The uncertainty in Fe$^{23+}$  collisional rate coefficients stem from the difference between the distorted-wave and R-matrix calculations for Fe$^{23+}$ \citep{2011A&A...528A..69L}. The uncertainties in dielectronic recombination  (DR) rates,  radiative recombination (RR) rates, and collisional ionization (CI) rates are taken from \citep{2021ApJ...908....3H}.}}
\begin{tabular}{c|c|c|c|c|c|c|c|c}
\hline
Ion &Label$^\dagger$ & Transition  & Energy (keV) & A error & q error & DR error  & RR error  & CI error \\
\hline
Fe$^{24+}$ &x &1s2p $^{3}\rm P_{2}\rightarrow  1s^{2}$ &6.6823&1\%&6\%&- &- &-\\
Fe$^{24+}$ &y &  1s2p $^{3}\rm P_{1}\rightarrow  1s^{2}$ &6.6675&14\%&18\%&- &- &-\\
Fe$^{24+}$ &z &1s2s $^{3}\rm S_{1}\rightarrow  1s^{2}$&6.6366&7\%&42\%&- &- &-\\
Fe$^{24+}$ &w &1s2p $^{1}\rm P_{1}\rightarrow  1s^{2}$&6.7004&2\%&7\%&- &- &-\\
Fe$^{23+}$ &s & 1s2s2p $^{3}\rm P_{3/2}\rightarrow 1s^{2}2s $ ($^{2}\rm S_{1/2}$) & 6.6619 &25\%&20\%&- &- &-\\
Fe$^{23+}$ &r & 1s2s2p $^{1}\rm P_{1/2}\rightarrow 1s^{2}2s $ ($^{2}\rm S_{1/2}$) & 6.6766&25\% &20\%&- &- &-\\
Fe$^{23+}$ &t & 1s2s2p $^{3}\rm P_{1/2}\rightarrow 1s^{2}2s $ ($^{2}\rm S_{1/2}$)& 6.6533&25\%&20\%&- &- &-\\
Fe$^{24+}$ & - & All transitions & -&-&-&20\% &20\% &15\%\\
Fe$^{23+}$ & -& All transitions & -&-&-&20\% &20\% &13\%\\
Fe$^{25+}$ &- & All transitions & -&-&-&20\% &20\% &15\% \\
\hline
\footnote{$^\dagger$ Labeled by Gabriel (1972)}
\end{tabular}
\label{t:c1}
\end{table*}

\section{The new AtomDb class \texttt{variableapec}}

\begin{figure*}
\gridline{\fig{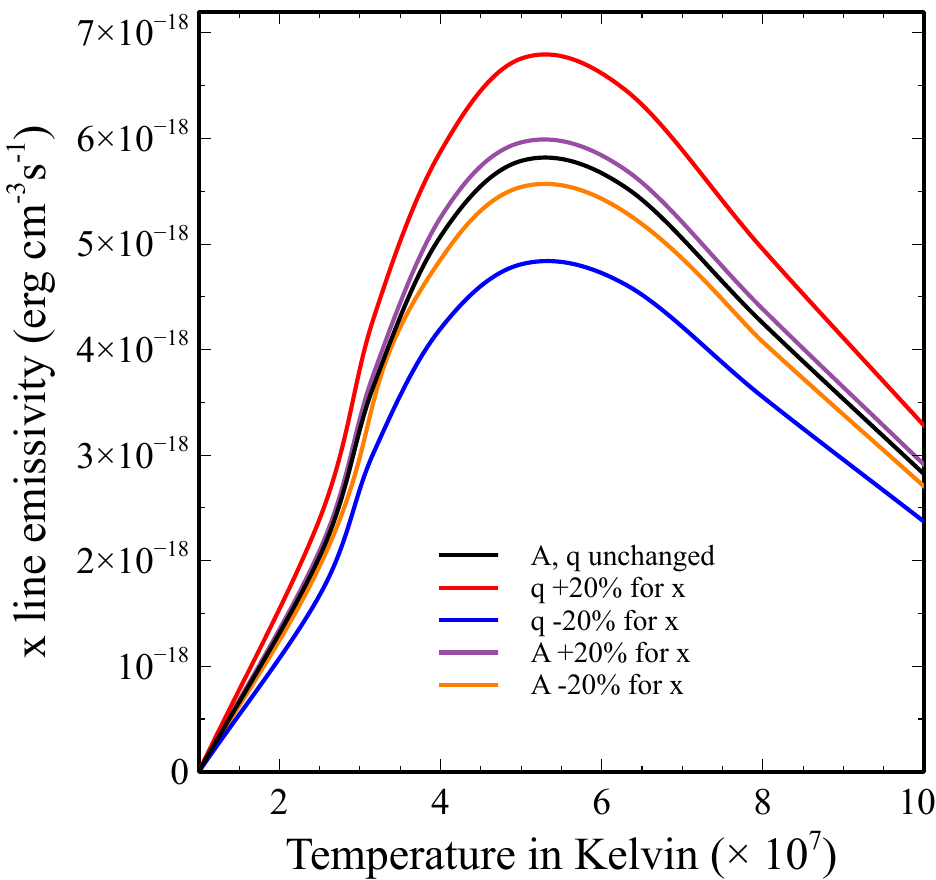}{0.32\textwidth}{(a)}
          \fig{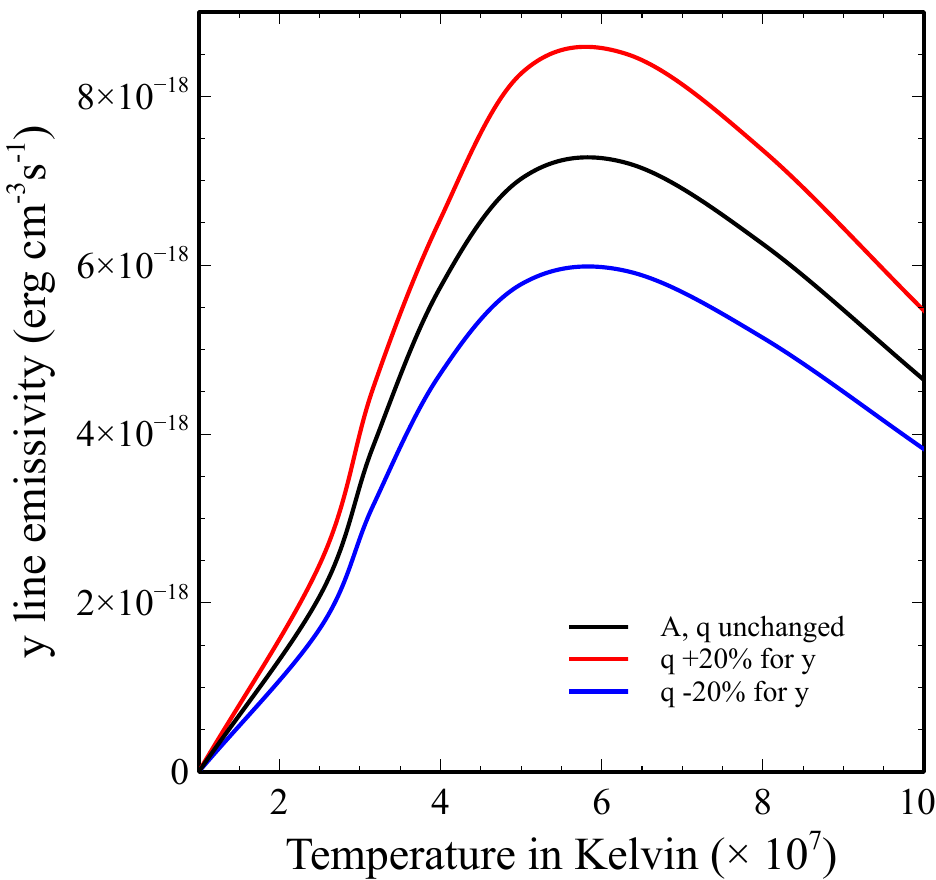}{0.32\textwidth}{(b)}
          \fig{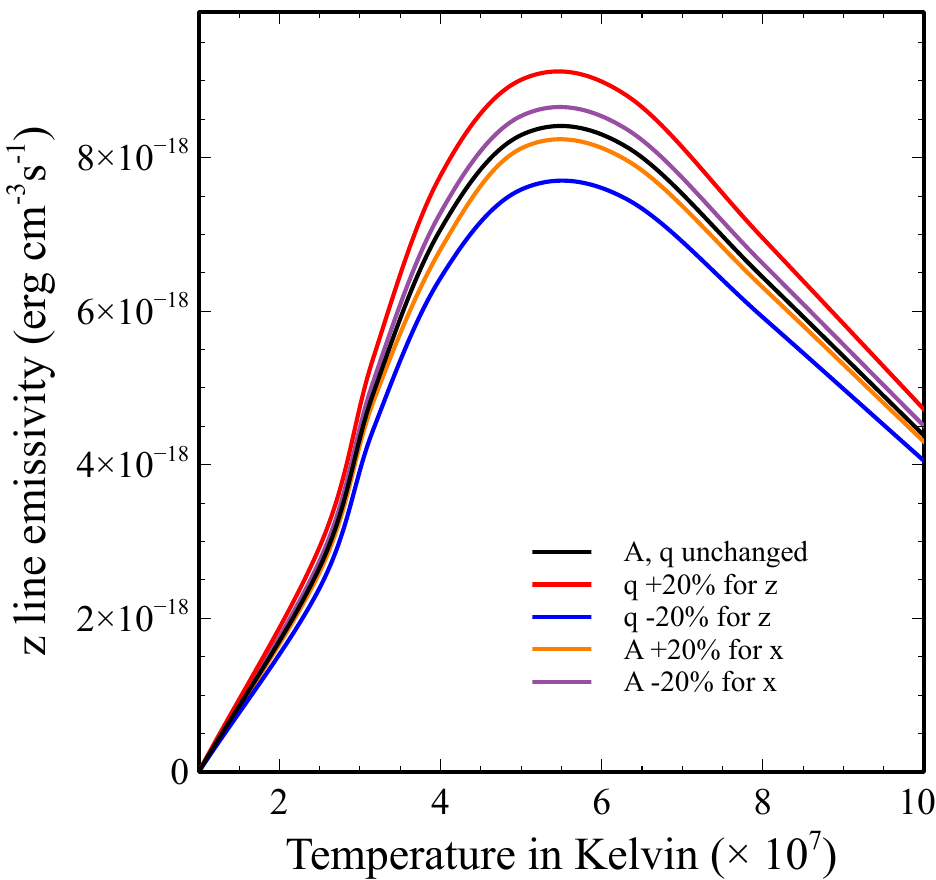}{0.32\textwidth}{(c)}                 
          }
\gridline{\fig{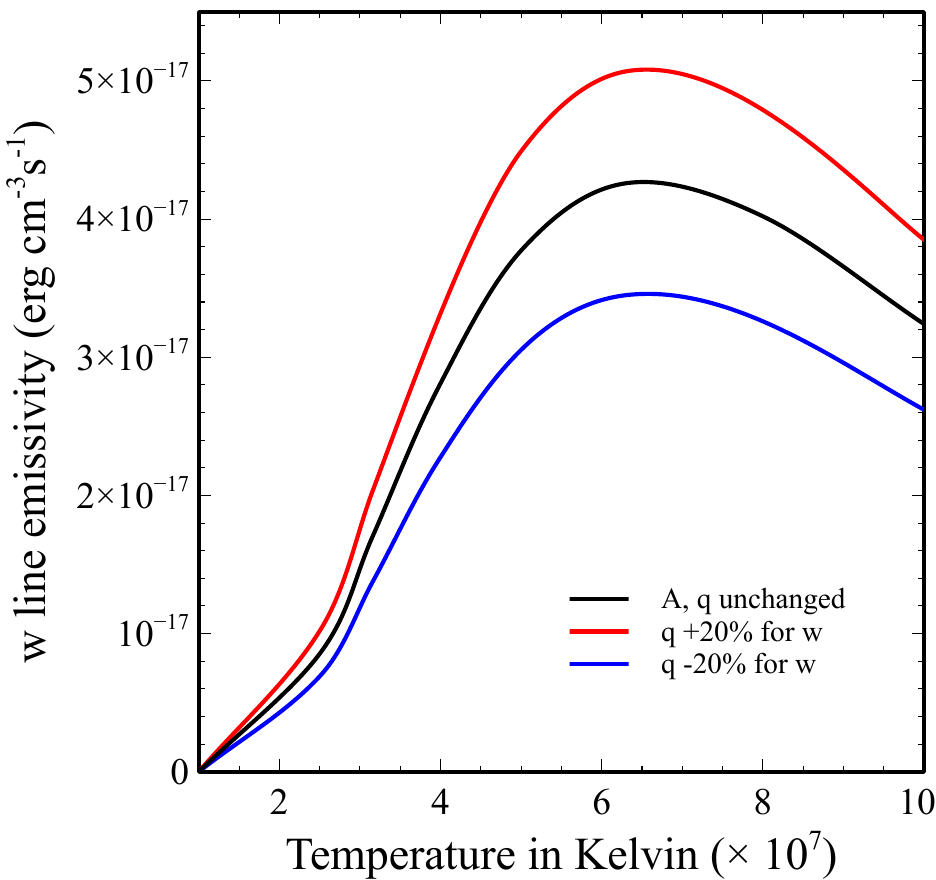}{0.32\textwidth}{(e)}
          \fig{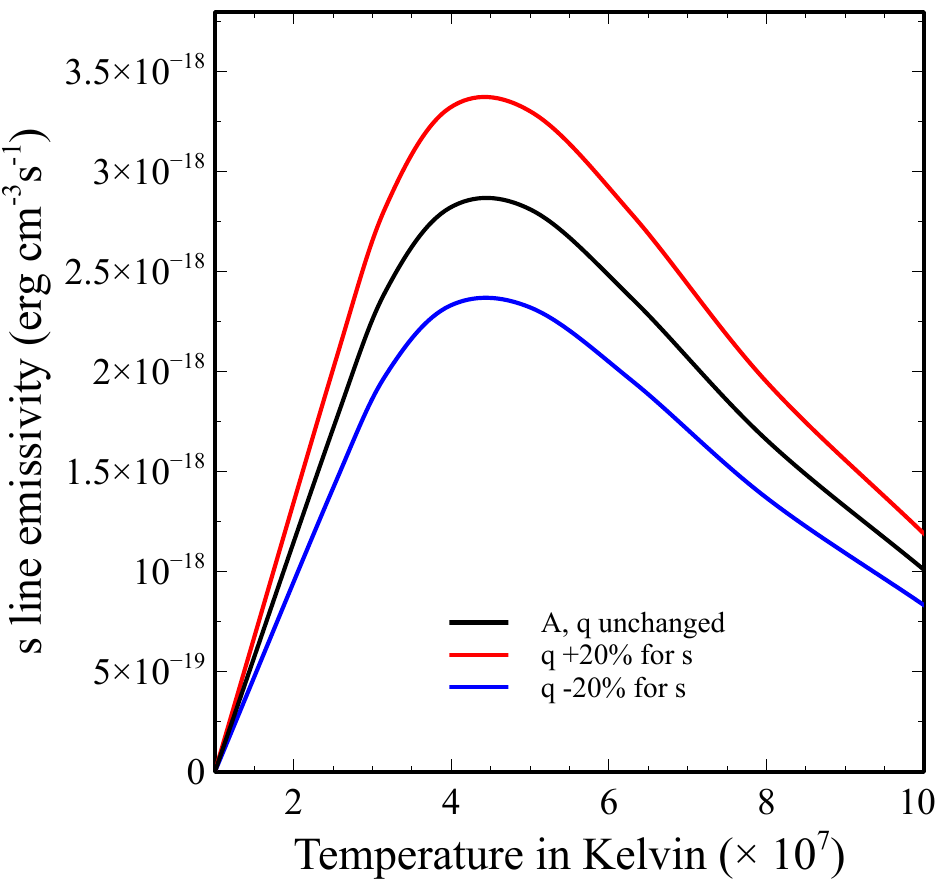}{0.32\textwidth}{(f)}
          \fig{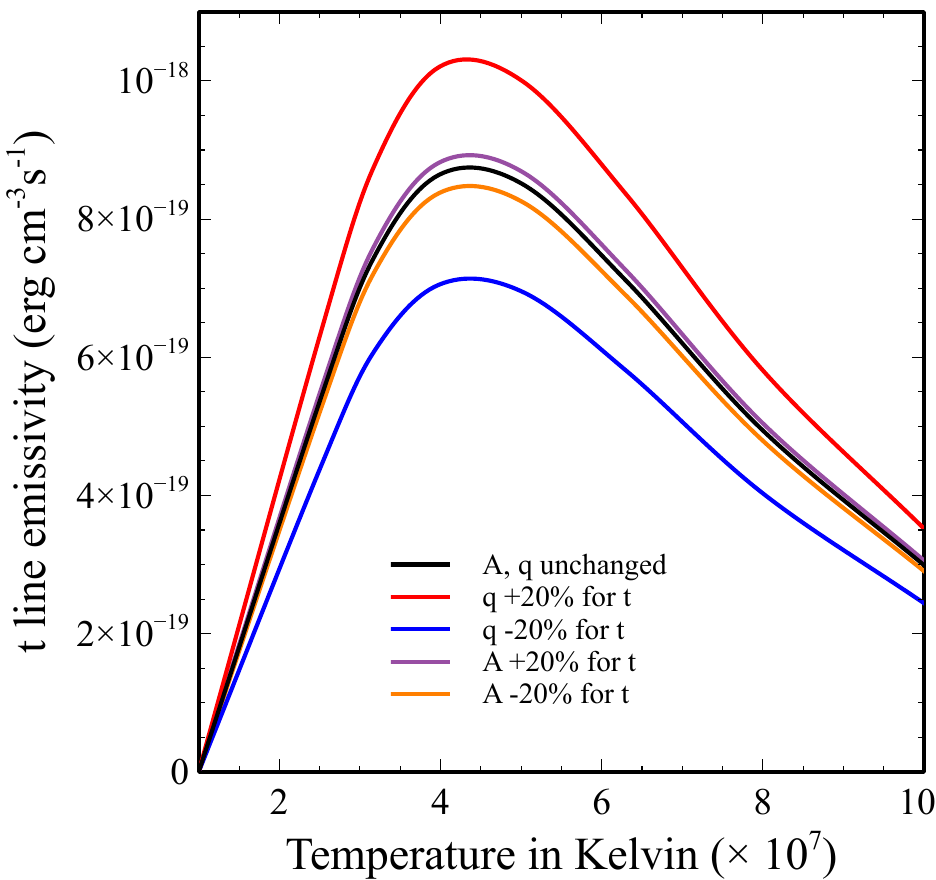}{0.32\textwidth}{(g)}
           }
\caption{Line emissivities of  x, y, z, w,  s, and t lines, with the   collisional rate coefficients (q) and Einstein A coefficients (A) varied, computed using \texttt{variableapec}, for electron density of 1 cm$^{-3}$.}
\label{fig:emiss}
\end{figure*}

There have been some recent changes in the infrastructure in AtomDB, which now has a class interface. The advantage of this interface is the flexibility to report the line emissivities 
just by specifying temperature, density, atomic number, and ion. One such class is 
 \texttt{variableapec}. The advantage of \texttt{variableapec} among the commonly used spectral fitting packages is that \texttt{variableapec} lets the users vary individual atomic rates such as Einstein A coefficients, electron  collisional rate coefficients, proton  collisional rate coefficients, and ionization/recombination rates for individual atomic transitions. It further facilitates reporting the resultant alterations in line emissivities, contributing to spectrum generation. The changed emissivities then can be fed into XSPEC to report the change in the best-fit fitting parameters, which has been further elaborated in Section \ref{fitting}.
 For a visual demonstration of the applicability of  \texttt{variableapec}, we vary the collisional rate coefficient (q values) and  Einstein A coefficients  (A values) for all the lines listed in table \ref{t:c1}. Line emissivities were calculated within the temperature range $10^7$K - $10^8$K and varying the q and A values by 20\% separately (see Figure \ref{fig:emiss}).  
 Varying  q values by 20\% varied the x, y, z, w, s, and t emissivities up to $\sim$ 20\%, $\sim$ 23\%,  $\sim$ 14\%,  $\sim$ 24\%, $\sim$ 21\%, and $\sim$ 22\%, respectively. Varying q by 20\% for the z line increases its emissivity by 14\% only, as there are other ways of populating the level 1s2s $^{3}\rm S_{1}$, such as 1s2p $^{3}\rm P_{2}\rightarrow$ 1s2s $^{3}\rm S_{1}$ and 1s2p $^{3}\rm P_{0}\rightarrow$ 1s2s $^{3}\rm S_{1}$. Varying A values by 20\% of the y and w transitions had no effect on y and w emissivities, increasing/decreasing A values of the x transition increased/decreased x line emissivity by $\sim$ 4\%, and decreased/increased the z line emissivity by $\sim$ 3\% \footnote{These numerical experiments
apply to the low-density limit, which for Fe XXV implies densities
less than  10$^{16}$ cm$^{3}$. In the high-density limit, z-line emissivity will directly depend on z Einstein A coefficient rather than indirectly depending on x Einstein A coefficient.}.
As y and w transitions have large A- values (4.46 $\times$ 10$^{13}$ s$^{-1}$ and 4.67 $\times$ 10$^{14}$ s$^{-1}$ respectively), the branching ratios are already 1, thus changing A- values for these transitions effectively has no effect on emissivities. There are two competing decay paths in x (1s2p $^{3}\rm P_{2}\rightarrow  1s^{2}$, and 1s2p $^{3}\rm P_{2}\rightarrow$  1s2s $^{3}\rm S_{1}$), so increasing the x emissivity by 20\% increases the branching ratio for x transition (A$_{1s2p ^{3} \rm P_{2}\rightarrow  1s^{2}}$/A$_{1s2p ^{3} \rm P_{2}\rightarrow  1s^{2}}$ + A$_{1s2p ^{3} \rm P_{2}\rightarrow  1s2s ^{3} \rm S_{1}}$) by $\sim$ 4\% and consequently decreases the z emissivity.

\begin{figure}[h!!]
\centering
\includegraphics[scale=0.40]{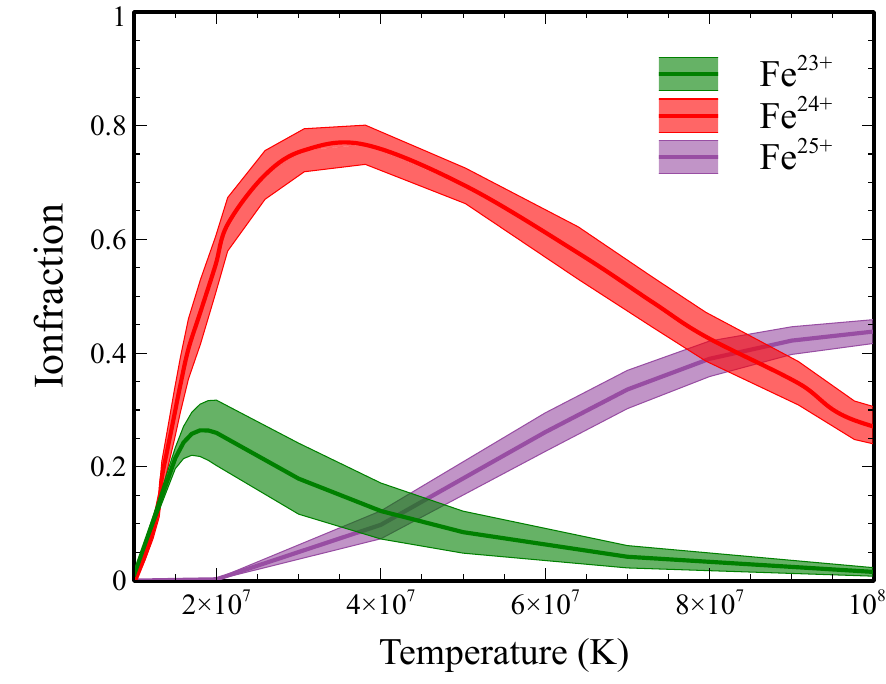}
\caption{Ion fraction of Fe$^{23+}$, Fe$^{24+}$, and Fe$^{25+}$. The shaded regions enclose the error in each ion fraction due to uncertainties in dielectronic recombination, radiative recombination, and collisional ionization rate listed in table \ref{t:c1}. }
\label{fig:if}
\end{figure}

The line emissivities are also dependent on the ion fraction uncertainties  of Fe$^{23+}$, Fe$^{24+}$, and Fe$^{25+}$. Figure \ref{fig:if} shows the variation in the ion fractions of the mentioned ions with temperature. The shaded regions enclose the uncertainties in the estimated ion fractions stemming from the error in  dielectronic recombination, radiative recombination, and collisional ionization rates in each of the ions. The enclosed ion fractions with uncertainties are taken from \citet{2021ApJ...908....3H}, who randomly sampled values within the uncertainty bounds of DR, RR, and CI errors and utilized Monte Carlo calculations to derive the range of potential ion fractions for a range of plasma temperatures.

\section{How much do the measured parameters vary?}

\subsection{Fitting parameters for no uncertainty in atomic data}\label{nochange}
\begin{figure}[h!!]
\centering
\includegraphics[scale=0.3]{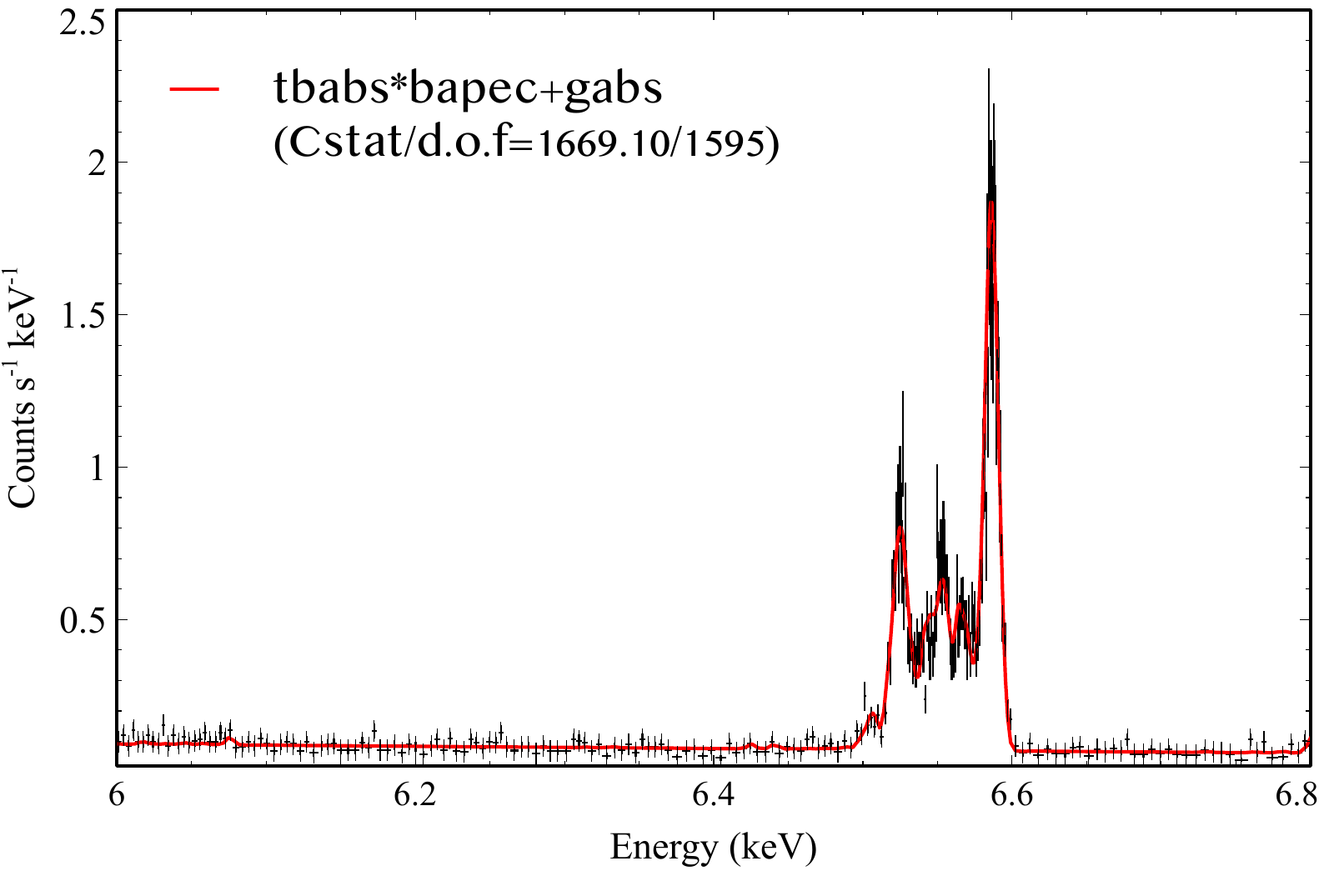}
\caption{X-ray spectrum of the Perseus cluster core fitted with m({=\tt\string tbabs*bapec+gabs}). }
\label{fig:perseus}
\end{figure}

We  started by fitting the observed spectrum within the energy range 6.0-6.8 keV, including all the Fe XXV and Fe XXIV emission lines with the pyXSPEC model m(=\texttt{tbabs*bapec+gabs}) with no uncertainty in the atomic data.   The routine {\tt tbabs}  accounts for the galactic absorption with the absorbing hydrogen column density set to 1.38$\times$10$^{21}$ cm$^{-2}$ \citep{2005A&A...440..775K} and  \texttt{gabs} accounts for the resonance scattering suppression in the w line as described in \citet{2018PASJ...70...10H}. The best-fit parameters for the fit are listed in table \ref{t:c2}. The spectrum of interest is displayed in Figure \ref{fig:perseus}.




\begin{figure}[h!!]
\centering
\includegraphics[scale=0.3]{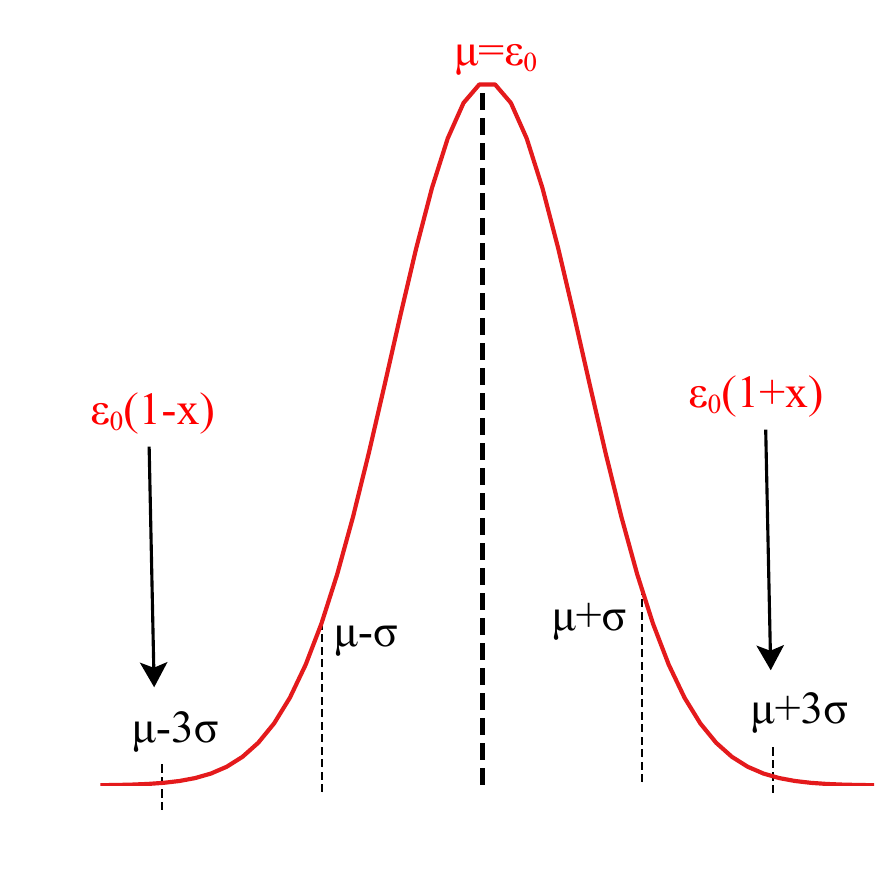}
\caption{Schematic representation of the Gaussian distribution of random numbers with a mean value of $\epsilon_{0}$ (line emissivity for zero uncertainty) and 3$\sigma$ value of $\pm$ $\epsilon_{0}$x , where x is the fractional value of the uncertainty. }
\label{fig:gauss}
\end{figure}

\begin{table*}
\centering{    
\caption{Best fit parameters for the observed \textit{Hitomi} spectrum of Perseus cluster using the model {\tt\string tbabs*bapec+gabs} within the energy range 6.0-6.8 keV. Solar abundance table by \citet{2009LanB...4B..712L} was used for the XSPEC fitting. All errors are reported within 3$\sigma$ interval. The redshift is taken from \citet{2020A&A...633A..42S} and has been frozen. \label{t:c2} }}
\begin{tabular}{c|c}
\hline
Parameters & Best-fit values \\
\hline
Temperature (keV) & $4.08^{+0.19}_{-0.18}$   \\
Iron Abundance & $0.71^{+0.07}_{-0.07}$  \\
Redshift (frozen) & 1.72284E-02 \\
Velocity Broadening (km/s) & $151.56 ^{+7.23}_{-7.11}$  \\
Norm (apec) & $0.79^{+0.09}_{-0.09}$ \\
Centroid of negative Gaussian (frozen, keV) & 6.587 \\
Sigma of negative Gaussian (keV)  & 3.992 $\times 10^{-3}$  \\
RS factor & $1.26^{+0.14}_{-0.14}$\\
C-stat/d.o.f & 1669.10/1595\\ 
\hline
\end{tabular}
\end{table*}

\subsection{Fitting parameters including uncertainty in atomic data}\label{fitting}

Although table \ref{t:c1} lists the maximum uncertainty values/percentages of errors in atomic rates, we could not find the distribution of the individual errors even after a rigorous search. Due to the lack of comprehensive information regarding the statistical distribution of uncertainties, we adopted a Gaussian distribution, a common practice in astronomy for model-based  parameter estimates \citep{2021RNAAS...5...39K, 2010arXiv1009.2755A}. 
To determine the sensitivity of the measured parameters, we varied the atomic data uncertainties for the listed lines in table \ref{t:c1} within their individual error ranges. For each line, we constructed a Gaussian distribution of 1000 random numbers with a  mean ($\mu$) of $\epsilon_{0}$, where $\epsilon_{0}$ is the line emissivity of an emission line with zero uncertainty.
We assumed the reported error values to be 3$\sigma$ values of the Gaussian to essentially cover the whole range of error under the Gaussian.
The widths of each Gaussian were set up such that if the fractional uncertainty is x, the lower-bound of the Gaussian ($\mu$-3$\sigma$) was set to be $\epsilon_{0}$(1-x), and upper bound ($\mu$+3$\sigma$) was set to be $\epsilon_{0}$(1+x), as shown in Figure \ref{fig:gauss}. This produced 1000 sets of line emissivities for each of the lines in the table.
Then we fitted the observed data with the 1000 sets of modified emissivities which were fed into XSPEC and fitted with the model \texttt{m}.

\begin{figure*}

\gridline{\fig{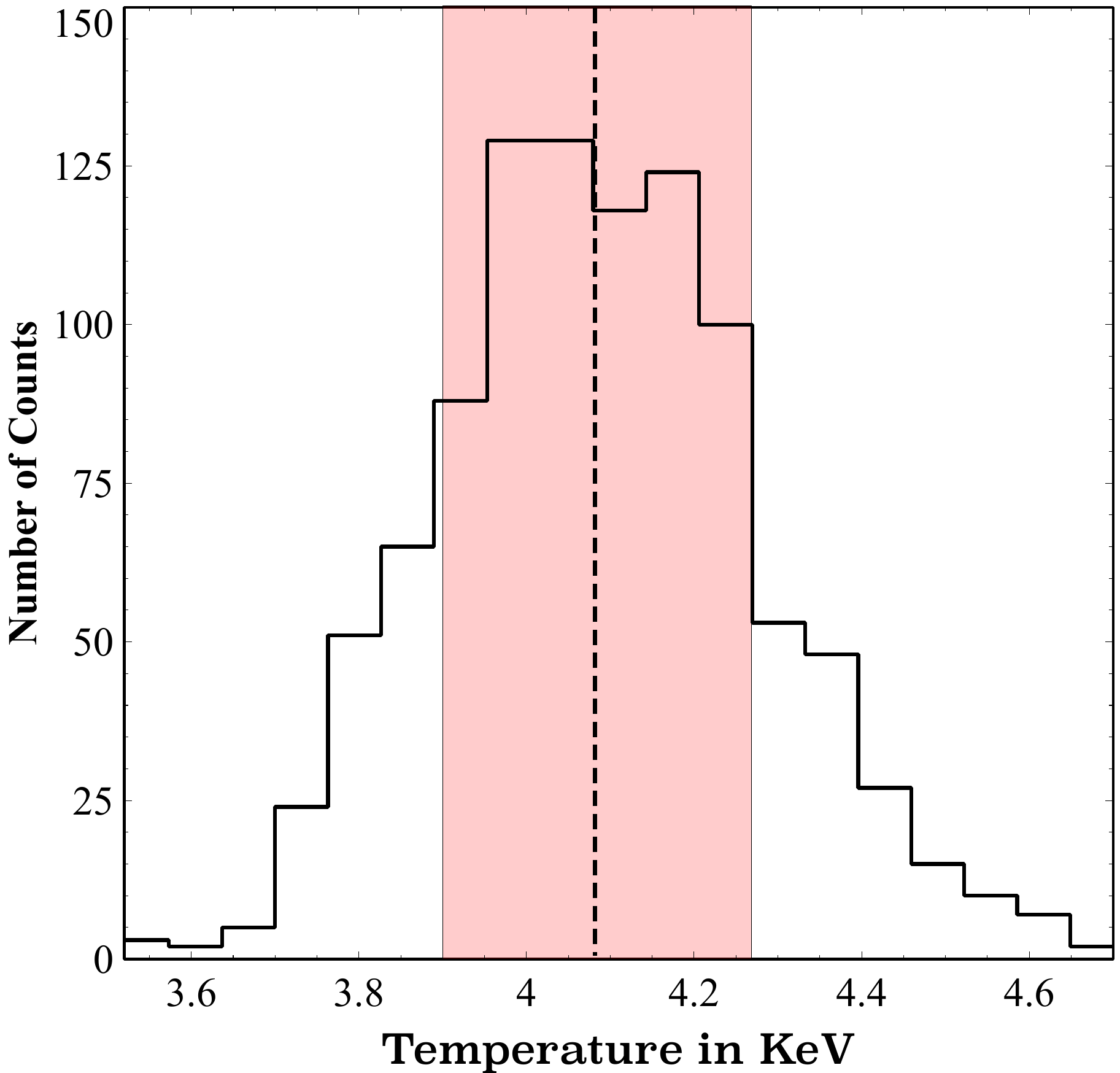}{0.25\textwidth}{(a)}
          \fig{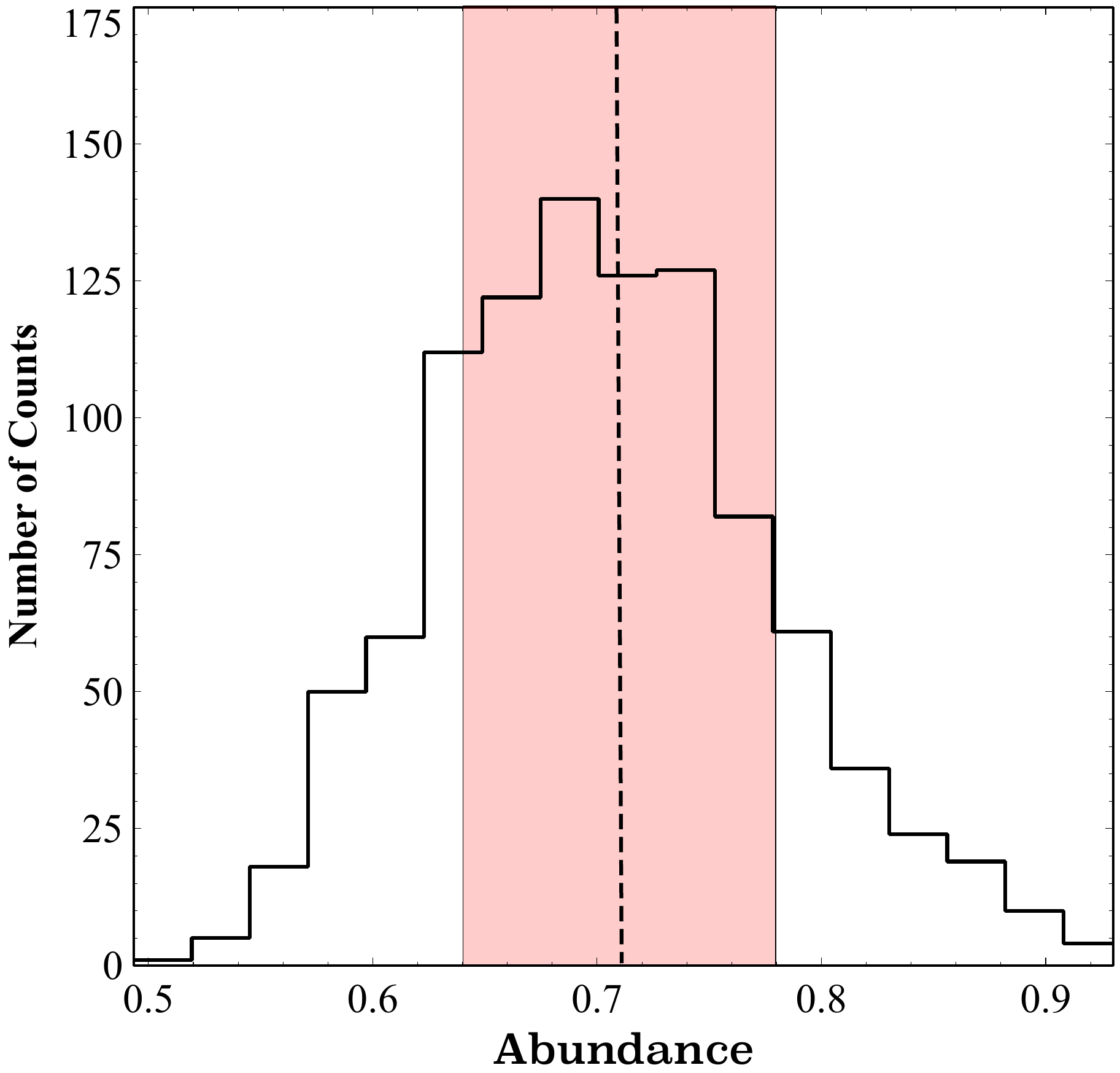}{0.25\textwidth}{(b)}
          \fig{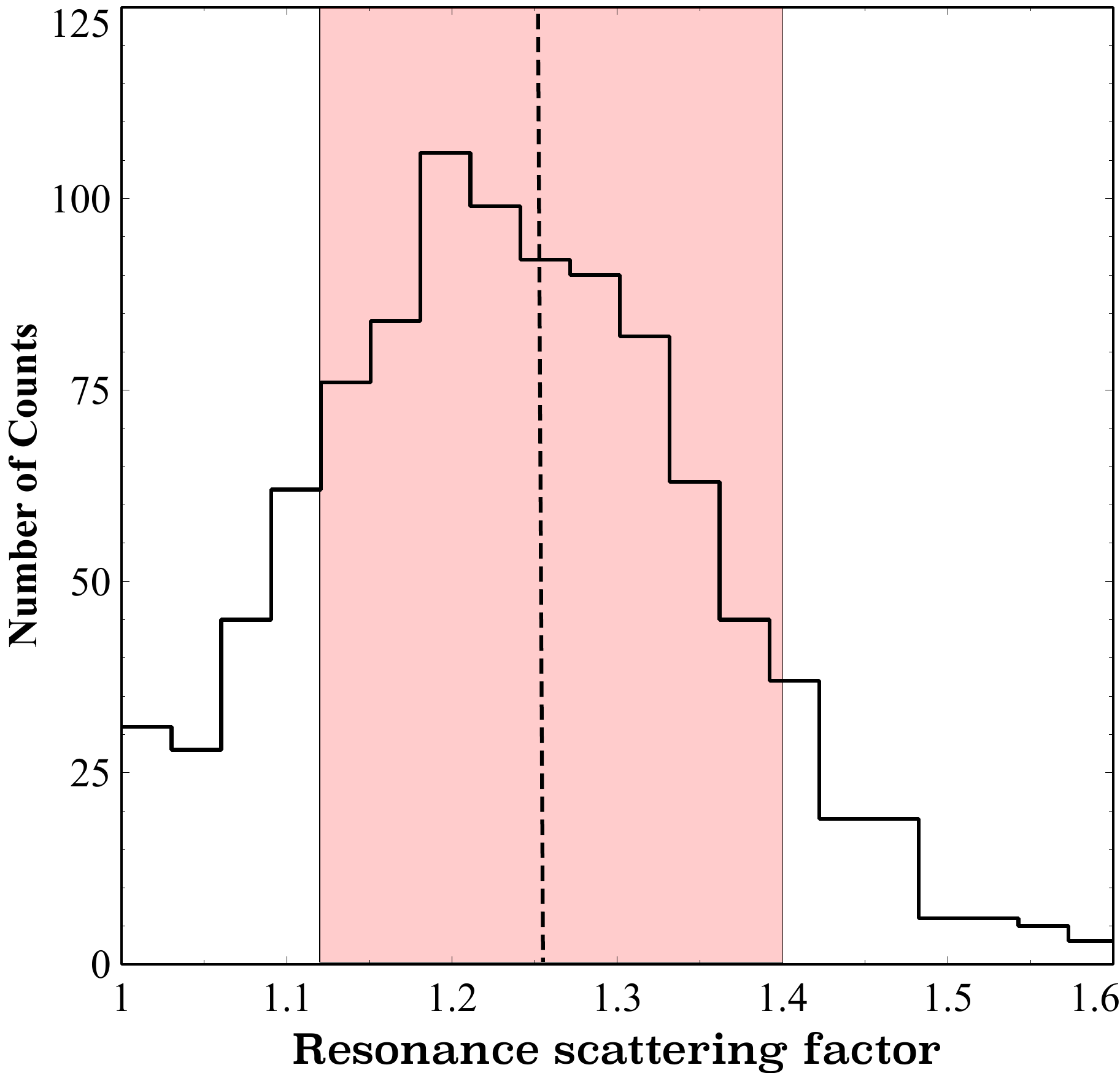}{0.25\textwidth}{(c)}
          \fig{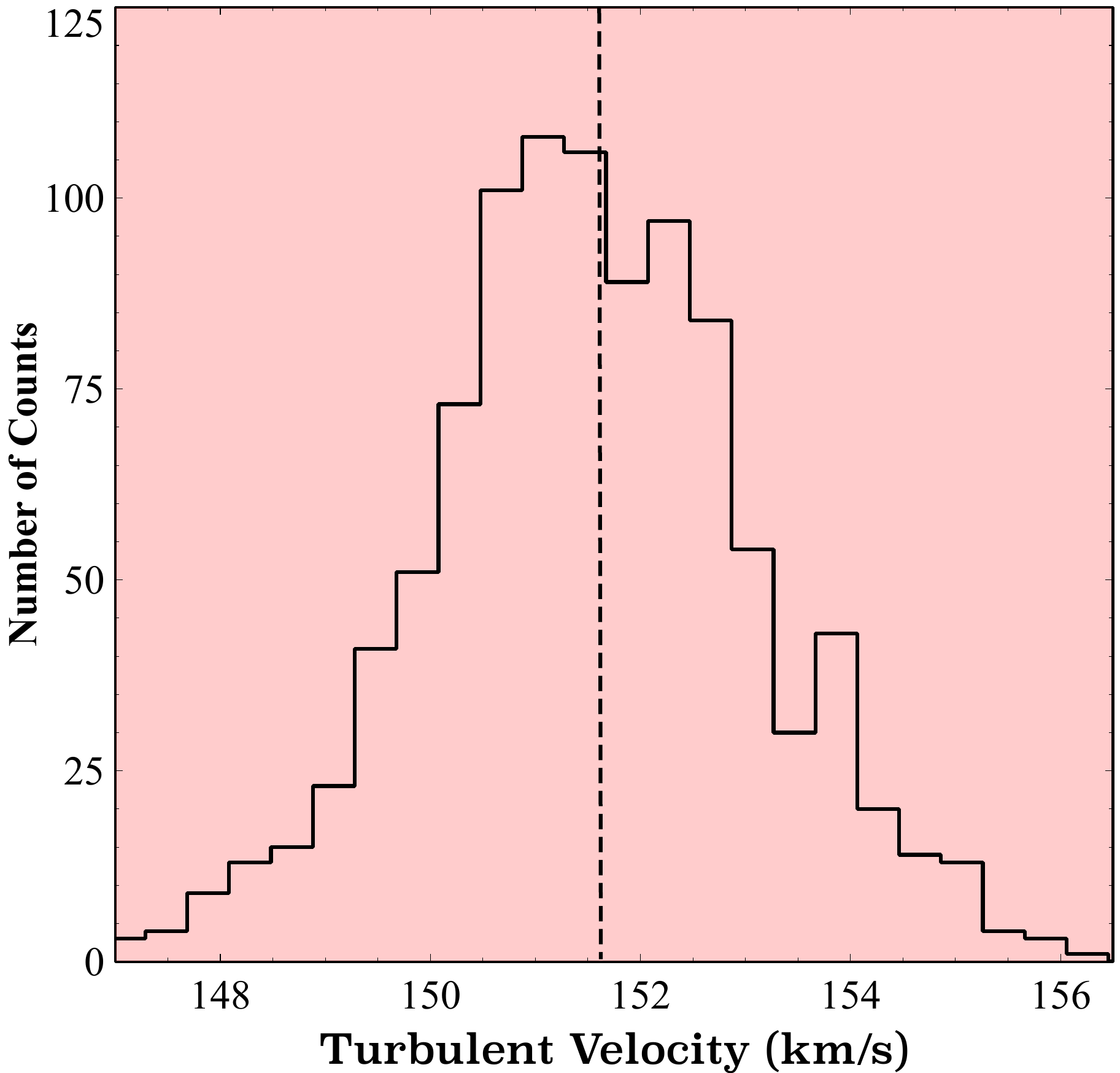}{0.25\textwidth}{(d)}
           }

\gridline{\fig{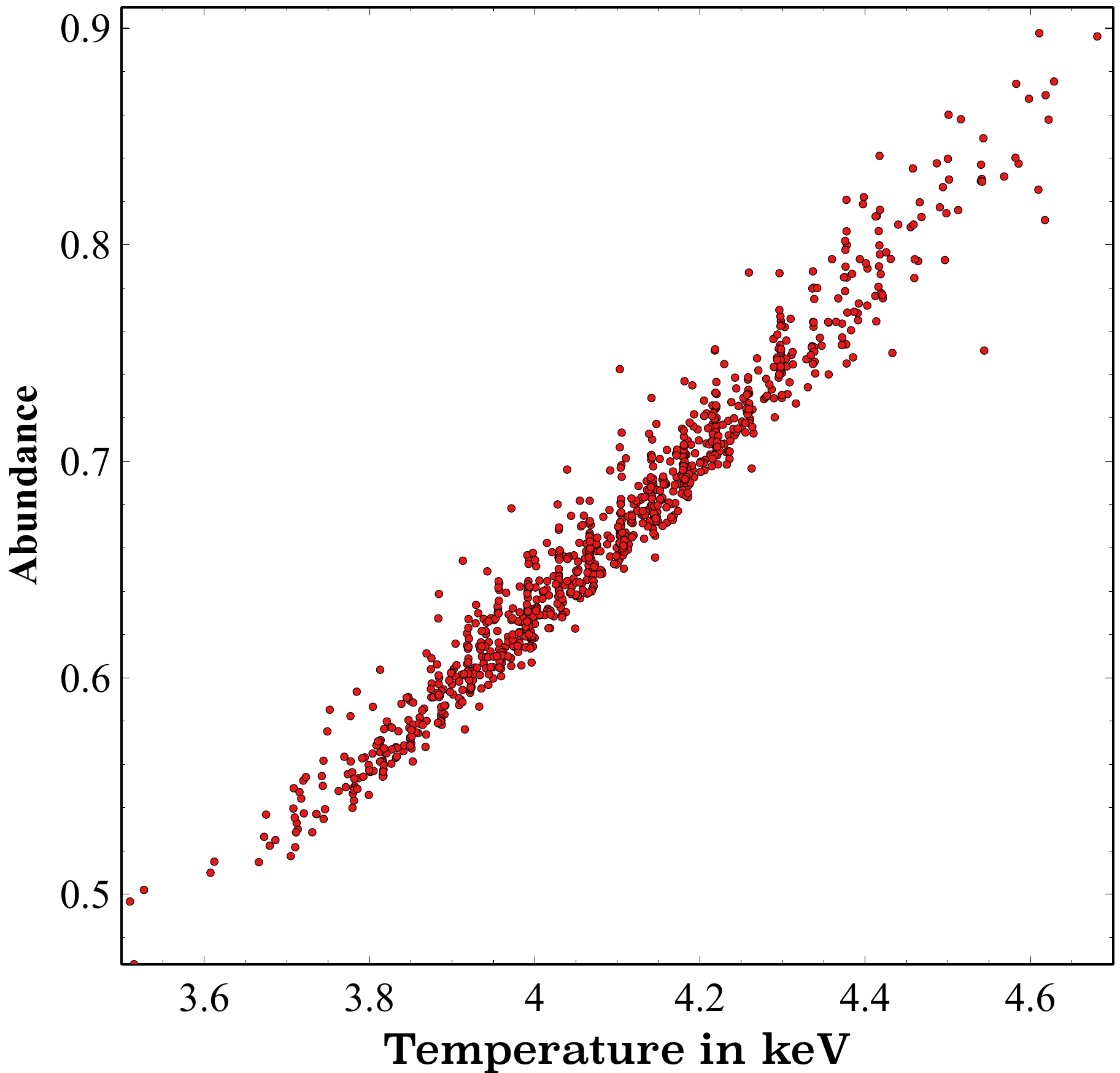}{0.25\textwidth}{(e)}
          \fig{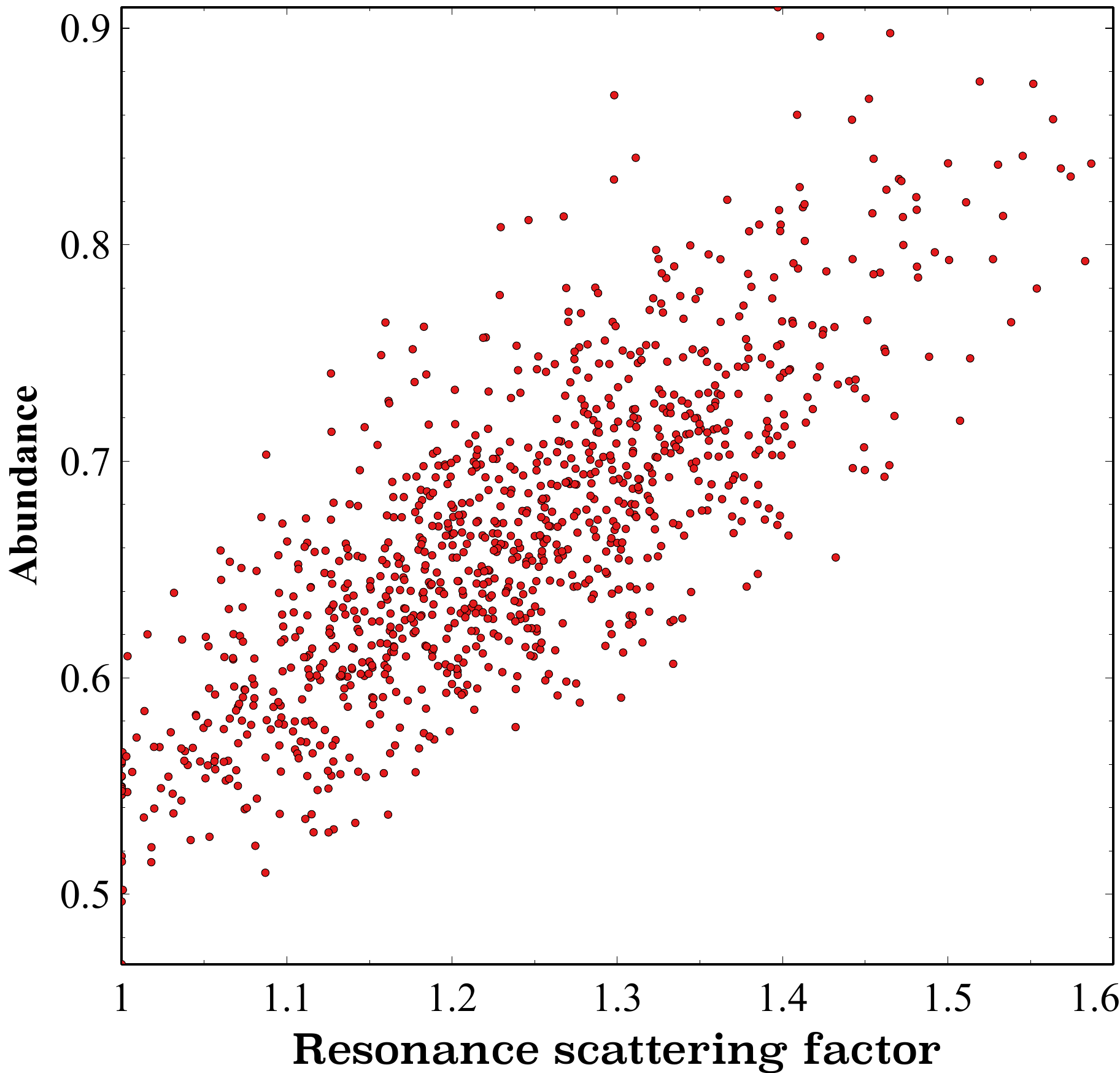}{0.25\textwidth}{(f)}
          \fig{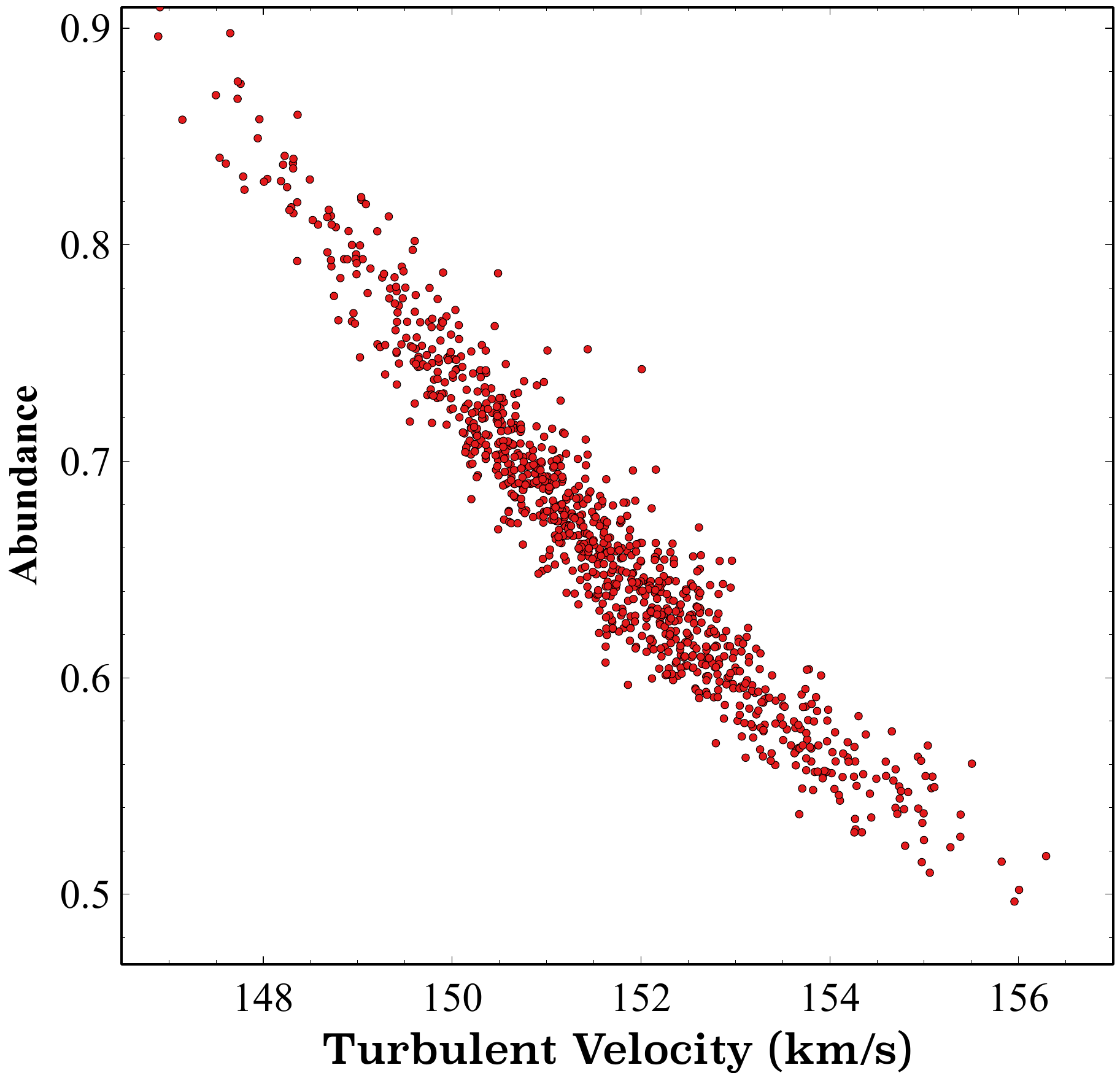}{0.25\textwidth}{(g)}         
          }
\caption{The top panels show  the distribution of temperature, abundance, resonance scattering factor, and  turbulent velocity for 1000 runs with varying Einstein A coefficient, collisional rate coefficient, ionization rate, dielectronic recombination, and radiative recombination uncertainties listed in table \ref{t:c1}. 
The vertical black dashed lines show the best-fit values of the above parameters with no uncertainty taken from table \ref{t:c2}. The red-shaded regions display the range of observational uncertainties. The bottom panels show scatter plots of the abundance versus temperature, abundance versus resonance scattering factor, and abundance versus turbulent velocity. }
\label{fig:3}
\end{figure*}

We found a variation in the best-fit temperatures within  3.51 to 4.71 keV. 
The best-fit abundance varied between 0.91 and 0.46. The measured resonance scattering factor varied from 1.0 to 1.64, and the turbulent velocity varied between 146.90 km/s and 156.30 km/s. 
The top panels of Figure \ref{fig:3} show   histograms  of the best-fit temperature, abundance, RS factor, and turbulent velocity, respectively. The red-shaded regions indicate their 3$\sigma$ error regions, implying that 32\%, 35\%, and 25\% of the best-fit
temperatures, abundances, and RS factors computed with the atomic data uncertainties lie outside the 3$\sigma$ error regions of their corresponding best-fit values computed with zero error in the atomic data. All the turbulent velocity values computed with atomic data uncertainties 
fall within the best-fit turbulence velocity 3$\sigma$ error range with no uncertainty in atomic data. The bottom panels of Figure \ref{fig:3} show  scatter plots of the abundance versus temperature, abundance versus resonance scattering factor, and abundance versus turbulent velocity. Note that using the observed Perseus core spectrum inherently leads to resonance suppression. We maintained the RS factor above 1 to ensure this.
However, in Section \ref{explore}, we explored the full parameter space and allowed the resonance scattering factor to dip below 1 to incorporate resonance enhancement.
The C-stat/d.o.f  values, obtained from 1000 fits, ranged from 1.04 to 1.12. We did not observe any correlation between the C-stat/d.o.f  values and the parameters mentioned above in our analysis, i.e., we were not making the fits significantly worse.

\begin{figure*}
\gridline{\fig{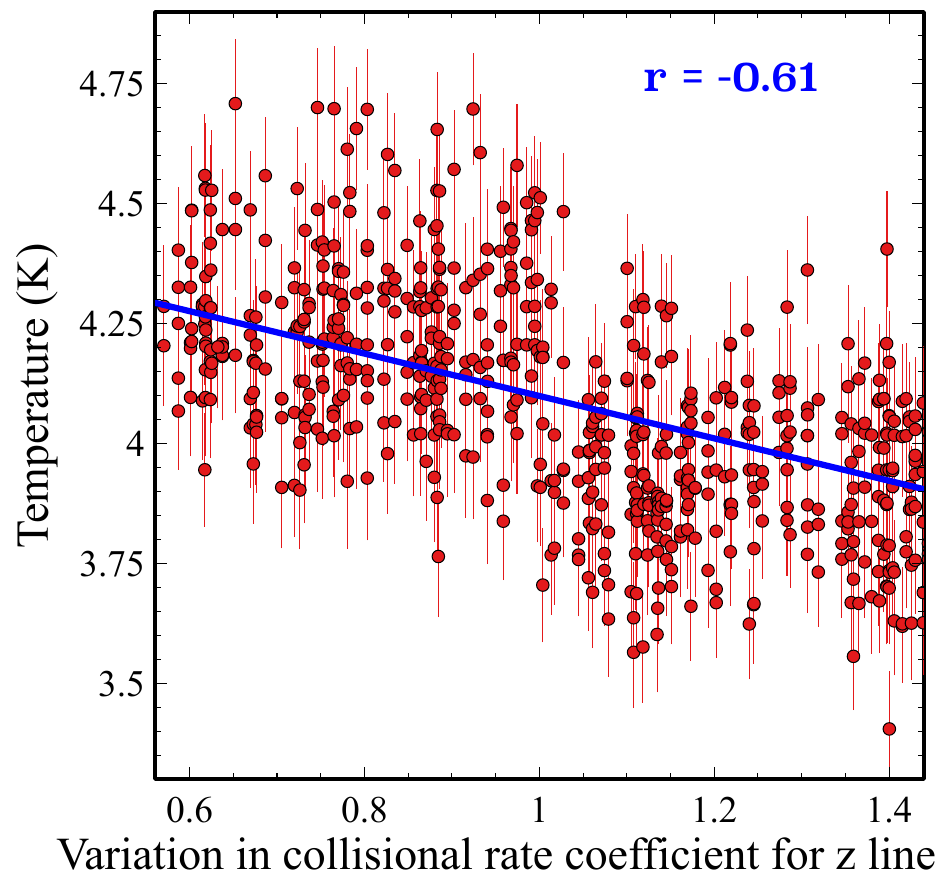}{0.3\textwidth}{(a)}
          \fig{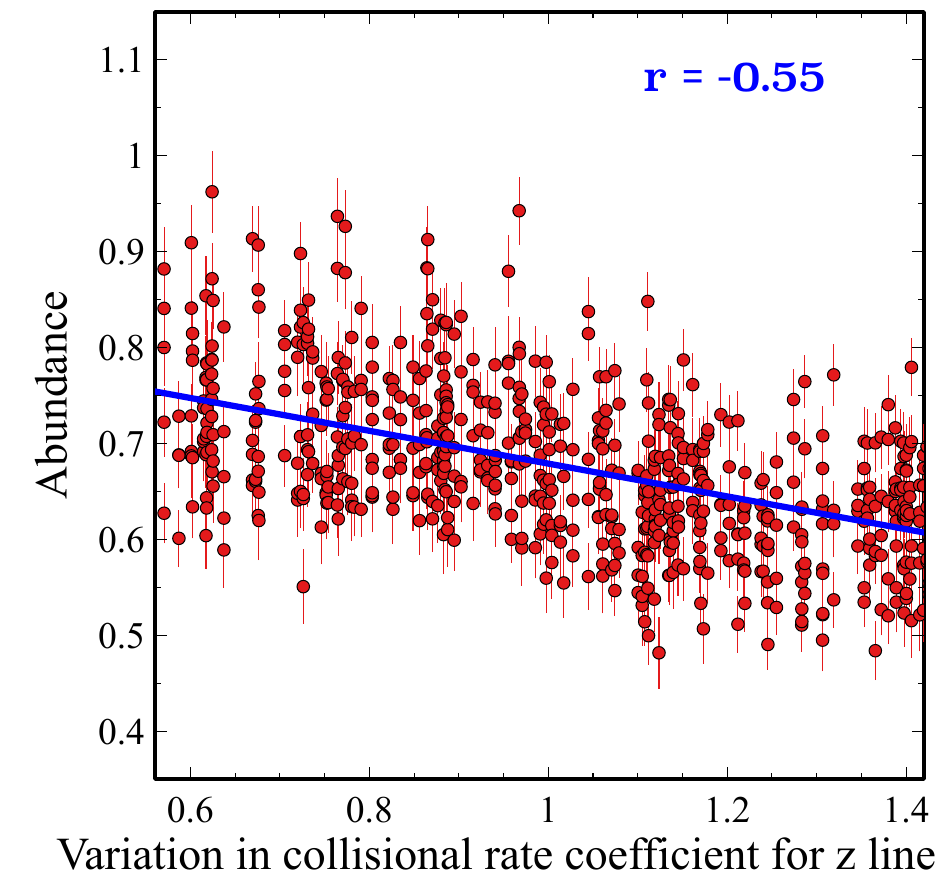}{0.3\textwidth}{(b)}
          \fig{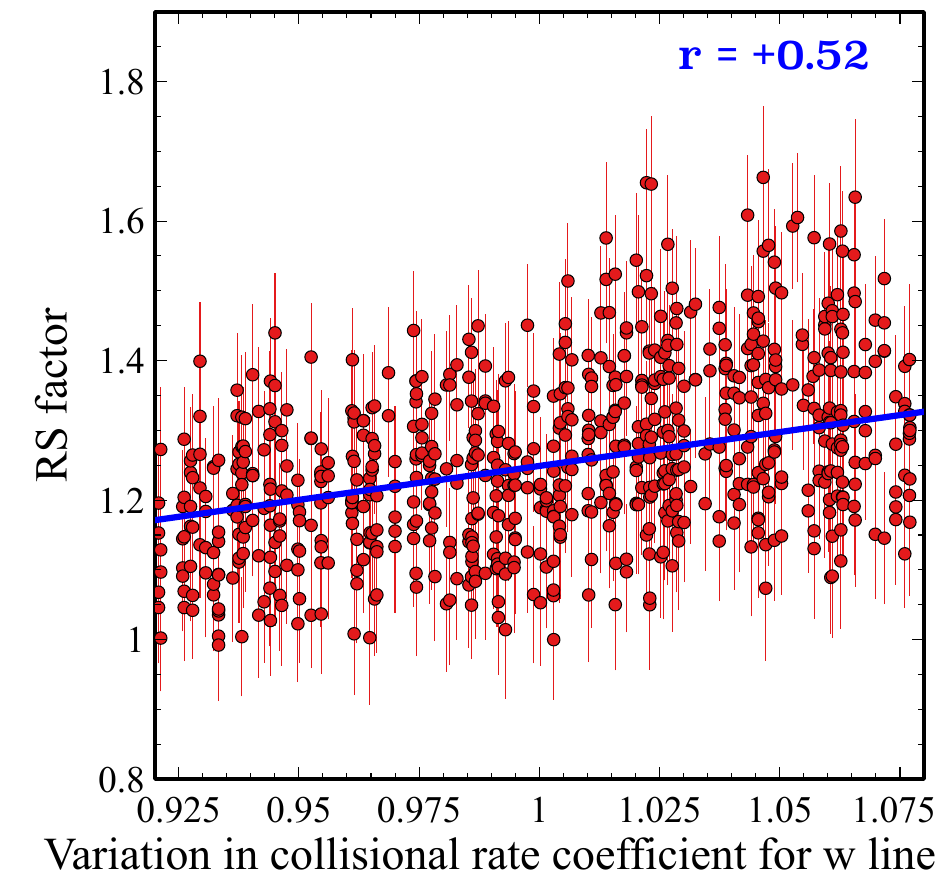}{0.3\textwidth}{(c)}
          }

\caption{Correlation between uncertainties and measured physical parameters for the actual atomic data uncertainties. The three panels show the correlation between temperature and z  collisional rate coefficient, abundance and z  collisional rate coefficient, resonance scattering factor, and w  collisional rate coefficient. }
\label{fig:1}
\end{figure*}

\section{Exploring the parameter space:}\label{explore}
The first step towards minimizing discrepancies between the data and the atomic models
is to determine the accuracy of which physical observable is correlated to the accuracy of which atomic dataset. We use the Pearson correlation coefficient for estimating the correlations between uncertainties and derived physical parameters \citep{freedman2007statistics}.  Pearson correlation coefficient is a statistical measure that assesses the linear relationship between two continuous variables. It measures the strength and direction of the relationship between the two variables, ranging from -1 (a perfect negative correlation) to +1 (a perfect positive correlation), with 0 indicating no correlation. The formula for Pearson correlation coefficient is: 
$r$ = $\frac{\sum xy - (\sum x)(\sum y)/n}{\sqrt{(\sum x^2 - (\sum x)^2/n)(\sum y^2 - (\sum y)^2/n)}}$, where $x$ and $y$ are the two variables being correlated, $\sum$ represents the sum of the values, $n$ is the number of data points, and $r$ is the correlation coefficient. We do not report correlations weaker than $r$=$\pm$ 0.5.

\begin{figure*}
\gridline{\fig{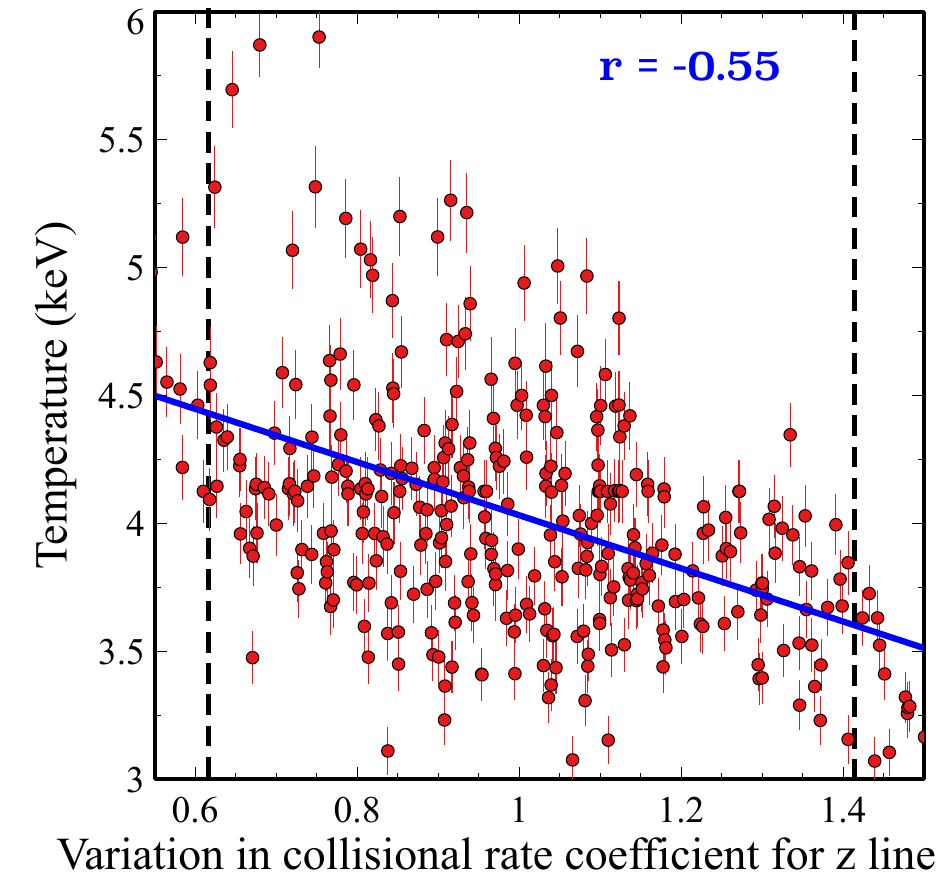}{0.35\textwidth}{(a)}
          \fig{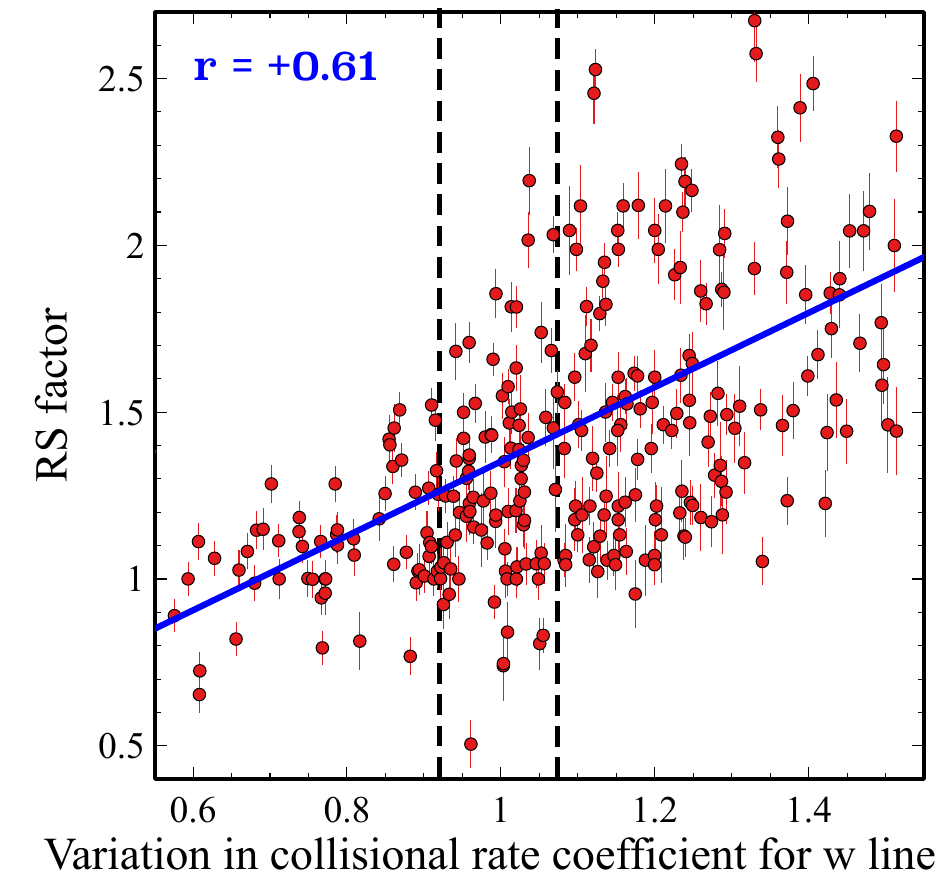}{0.35\textwidth}{(b)}
          }
\gridline{\fig{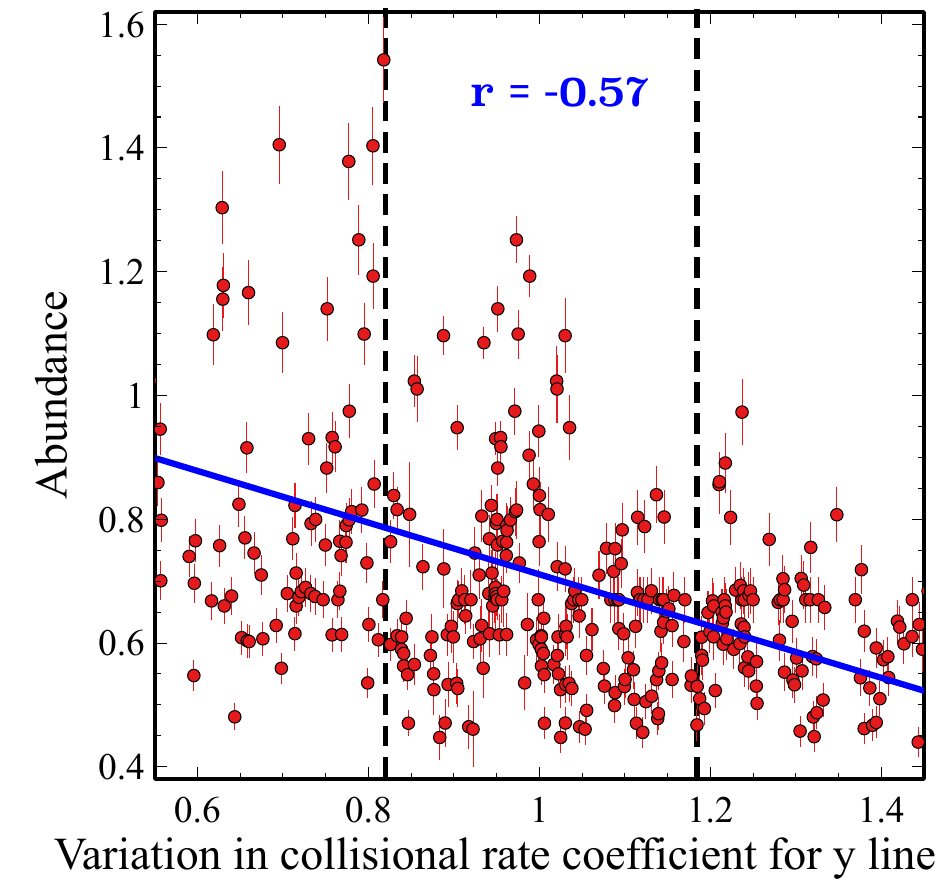}{0.35\textwidth}{(c)}
          \fig{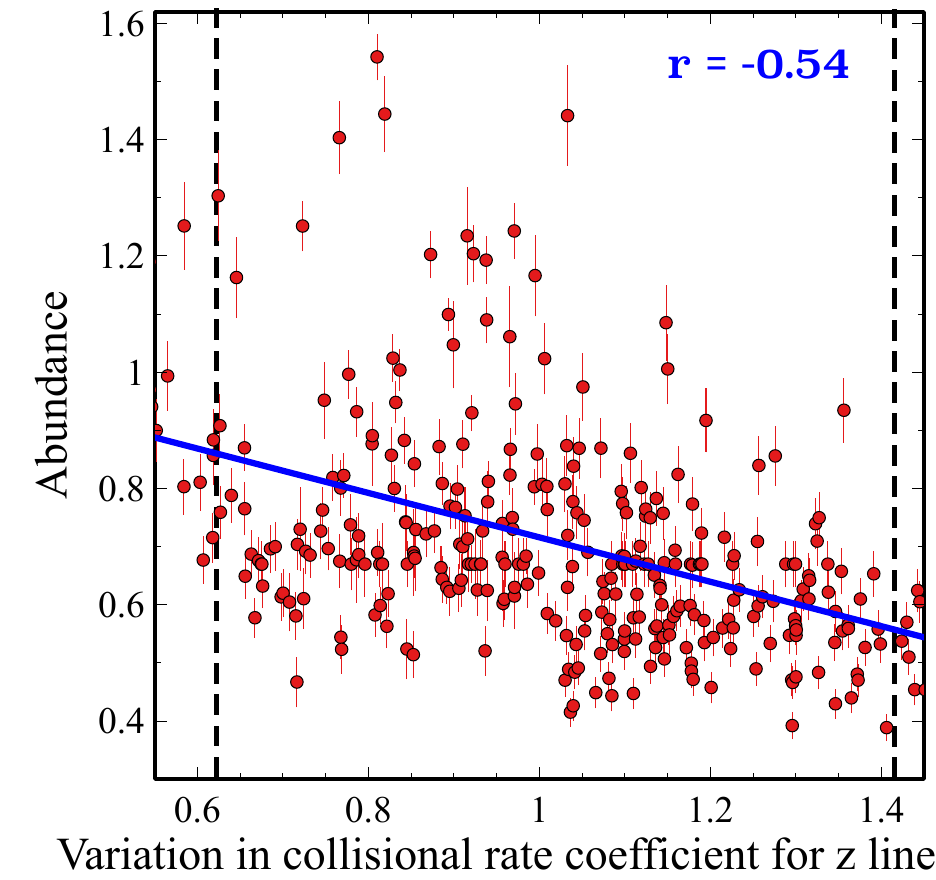}{0.35\textwidth}{(d)}
          }
\caption{Correlation between uncertainties and measured physical parameters varying all atomic data uncertainties by 45\%. The top three panels show the correlation between temperature and z Einstein A coefficient, temperature and z  collisional rate coefficient, RS factor and w   collisional rate coefficient. The two bottom panels display correlations between abundance and y  collisional rate coefficient and abundance and z  collisional rate coefficient. The dashed vertical lines represent the actual atomic data uncertainties listed in table \ref{t:c1}.}
\label{fig:corr2}
\end{figure*}

We started by finding correlations between Einstein A coefficients,  collisional rate coefficient, ionization/recombination rates with the measured temperature, metallicity, turbulence, and RS factor, varying the uncertainties with the percentages listed in table \ref{t:c1}. Figure \ref{fig:1} displays the correlations.
 We find a negative correlation with r=-0.61 between collisional rate coefficient in z and temperature and a weaker negative correlation between z  collisional rate coefficient and abundance with r=-0.55. A positive correlation with r=+0.52 was found between the RS factor and the  collisional rate coefficient of the w line. However, the correlations are highly biased towards the atomic datasets with the highest atomic data uncertainties. 
 
 
In order to obtain an unbiased measurement of the correlations, we need to apply the same uncertainty percentages to the atomic rates for all the lines. As the maximum fractional uncertainty listed in Table 1 is 0.42, we set the upper and lower limit of fractional uncertainty of each of the Fe XXV and Fe XXIV lines to be  1$\pm$0.45 to cover the entire uncertainty range for all the lines. Figure \ref{fig:corr2} displays the correlations.
The region between the dashed vertical lines in each of the plots represents the actual fraction of data uncertainties for each line. As expected, varying all uncertainties within the same percentage slightly weakened the correlation between z  collisional rate coefficient and temperature and z  collisional rate coefficient with abundance. We also found a new correlation between y  collisional rate coefficient and abundance. The correlation between the RS factor and w collisional rate coefficient strengthened. \\
We find  a weaker negative correlation between z  collisional rate coefficient and temperature with $r$ =- 0.55.  We find a positive correlation between the RS factor and the collisional rate coefficient in w with r=+0.59. This is expected as the w line is the only optically thick line and hence is susceptible to resonance scattering.
The bigger the collisional rate coefficient, the stronger the w line and the bigger the RS factor becomes to fit the line. We find negative correlations between the elemental abundance and y collisional rate coefficient and elemental abundance and z collisional rate coefficient with $r$ = -0.57 and $r$ = -0.54, respectively.
The emissivities being much weaker, no correlations were found for the atomic data uncertainties of the Fe XXIV lines. No significant correlations were found between turbulent velocity and Einstein A coefficient/ collisional rate coefficient of any of the lines. No correlation was found between any of the observable quantities with the ionization/recombination rates.

\section{Investigating the biases}

\subsection{Expanding the energy range}
Electron temperature can be obtained from line ratios, as explored above, or the shape of the broader continuum over a wider energy range. To examine how expanding the energy range affects the fitting process, we also conducted a fitting analysis on the identical spectrum within the 1.8-20.0 keV energy range. As anticipated, we observed a significant reduction in the error bars associated with the plasma parameters. The best-fit values for temperature, abundance, RS factor, and turbulence,  for the extended energy range of fitting, without factoring in uncertainties related to atomic data, were found to be $4.07^{+0.06}_{-0.06}$ keV, $0.73^{+0.02}_{-0.02}$, $1.25^{+0.11}_{-0.11}$, and  $150.79^{+5.55}_{-5.62}$ km/s, respectively.

We plotted the distribution of best-fit temperature, abundance, resonance scattering factor, and turbulence velocity within the 1.8-20.0 keV energy range based on 1000 runs, varying atomic data uncertainties as described in section \ref{fitting}.  We observed a temperature variation ranging from 4.02 to 4.13 keV in the best-fit values. The best-fit abundance showed a range of 0.67 to 0.78. The resonance scattering factor ranged from 1.0 to 1.53, and the turbulent velocity varied between 146.60 km/s and 156.50 km/s.
The top panel of Figure \ref{fig:20keV} displays the distributions, indicating that the temperatures and turbulence calculated with atomic data uncertainties remain within the 3$\sigma$ error intervals of their respective best-fit values computed with zero error in the atomic data.
This is anticipated, given that the key factor influencing the temperature estimate is the curvature of the Bremsstrahlung continuum, known to have a small uncertainty, which significantly narrowed the temperature distribution. The abundance distribution narrowed compared to that of the 6.0-6.8 keV range, with 26\% of the best-fit abundances with atomic data uncertainties falling outside the 3$\sigma$ error regions of their corresponding values with zero atomic data error. The RS factor distribution narrowed slightly,  with 22\%  of the best-fit values deviating beyond the 3$\sigma$ error regions compared to the without uncertainty error values.

\subsection{Multi-temperature plasma }

\begin{figure*}
\gridline{\fig{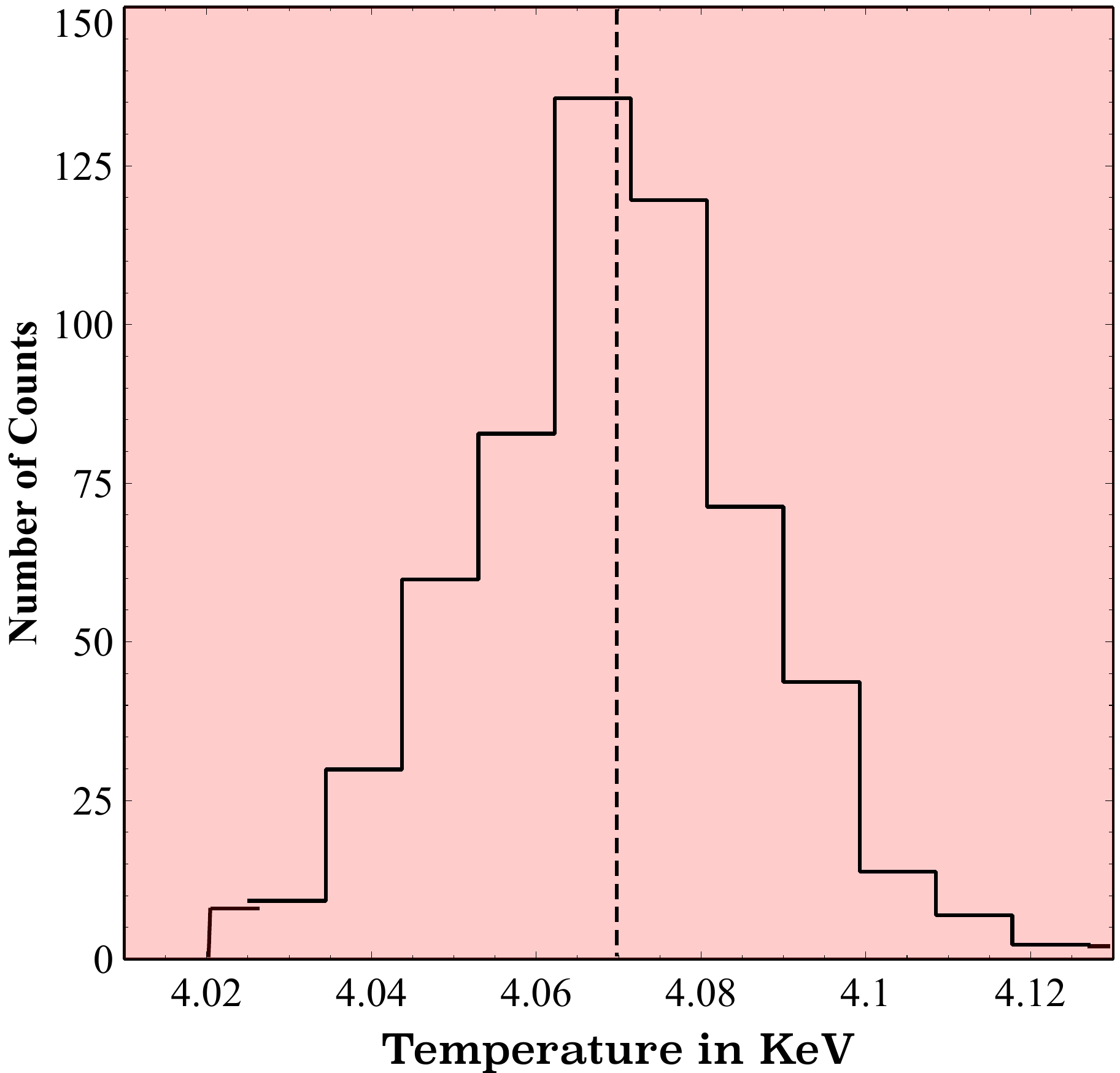}{0.25\textwidth}{(a)}
          \fig{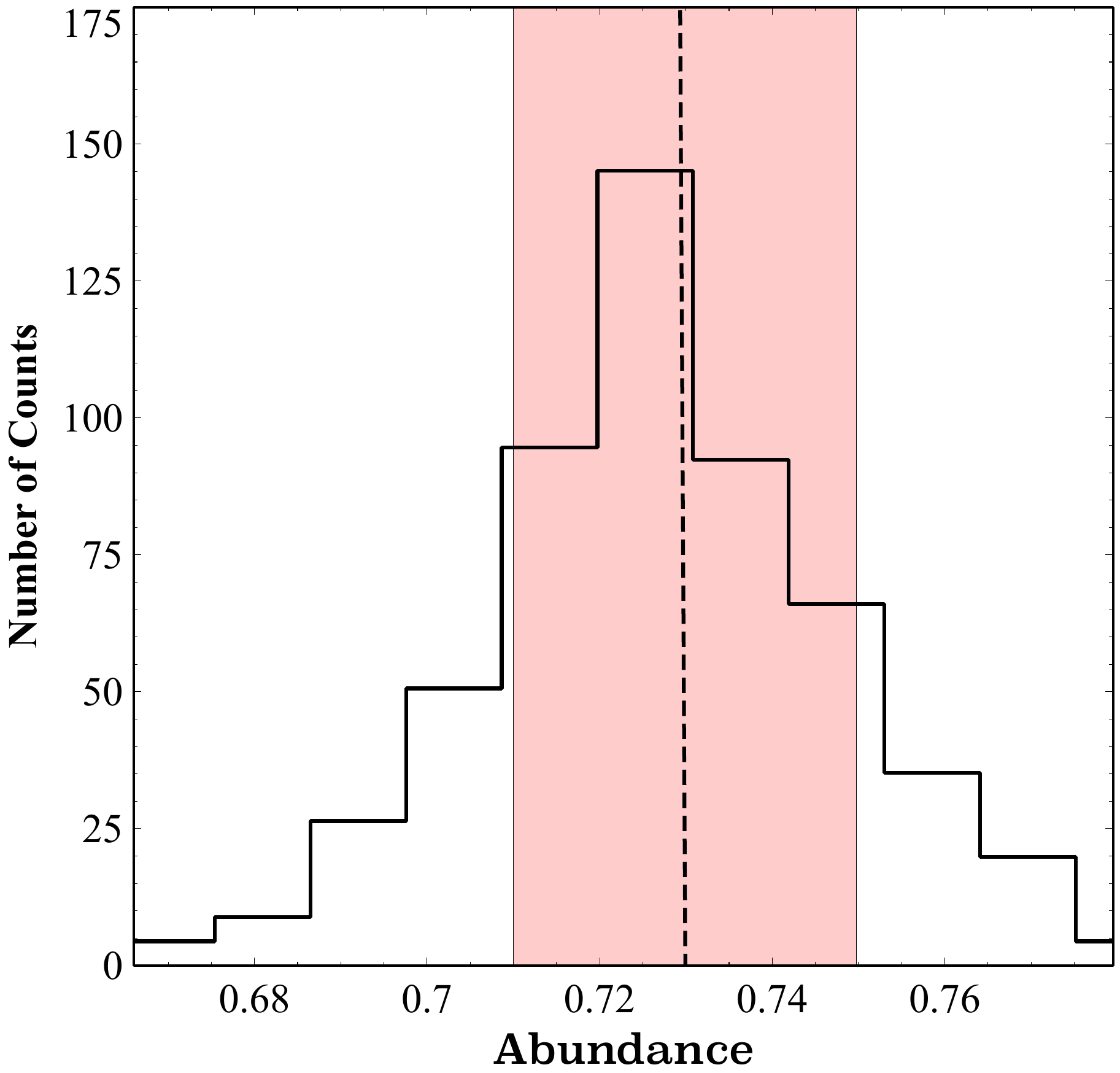}{0.25\textwidth}{(b)}
          \fig{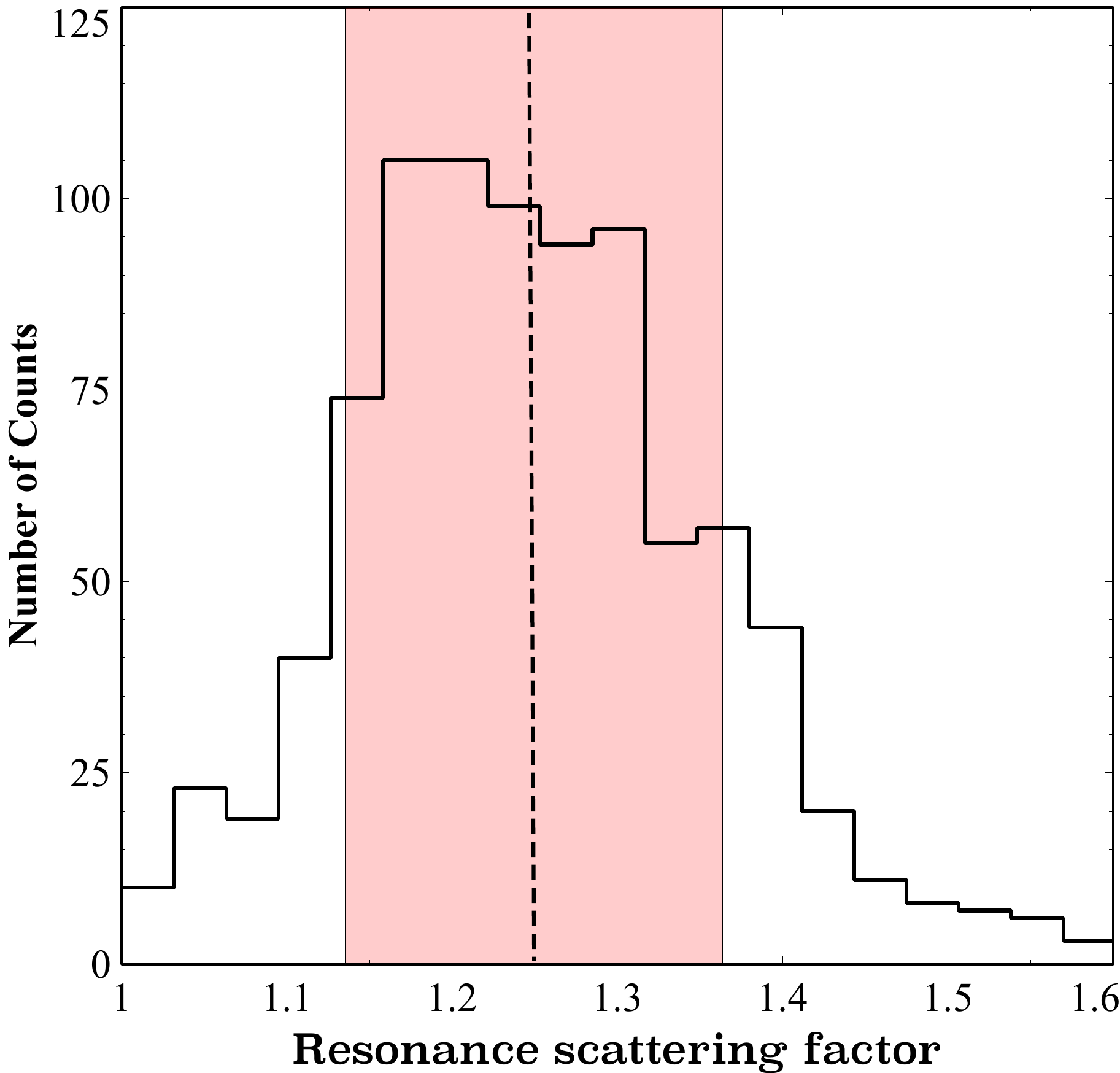}{0.25\textwidth}{(c)}
          \fig{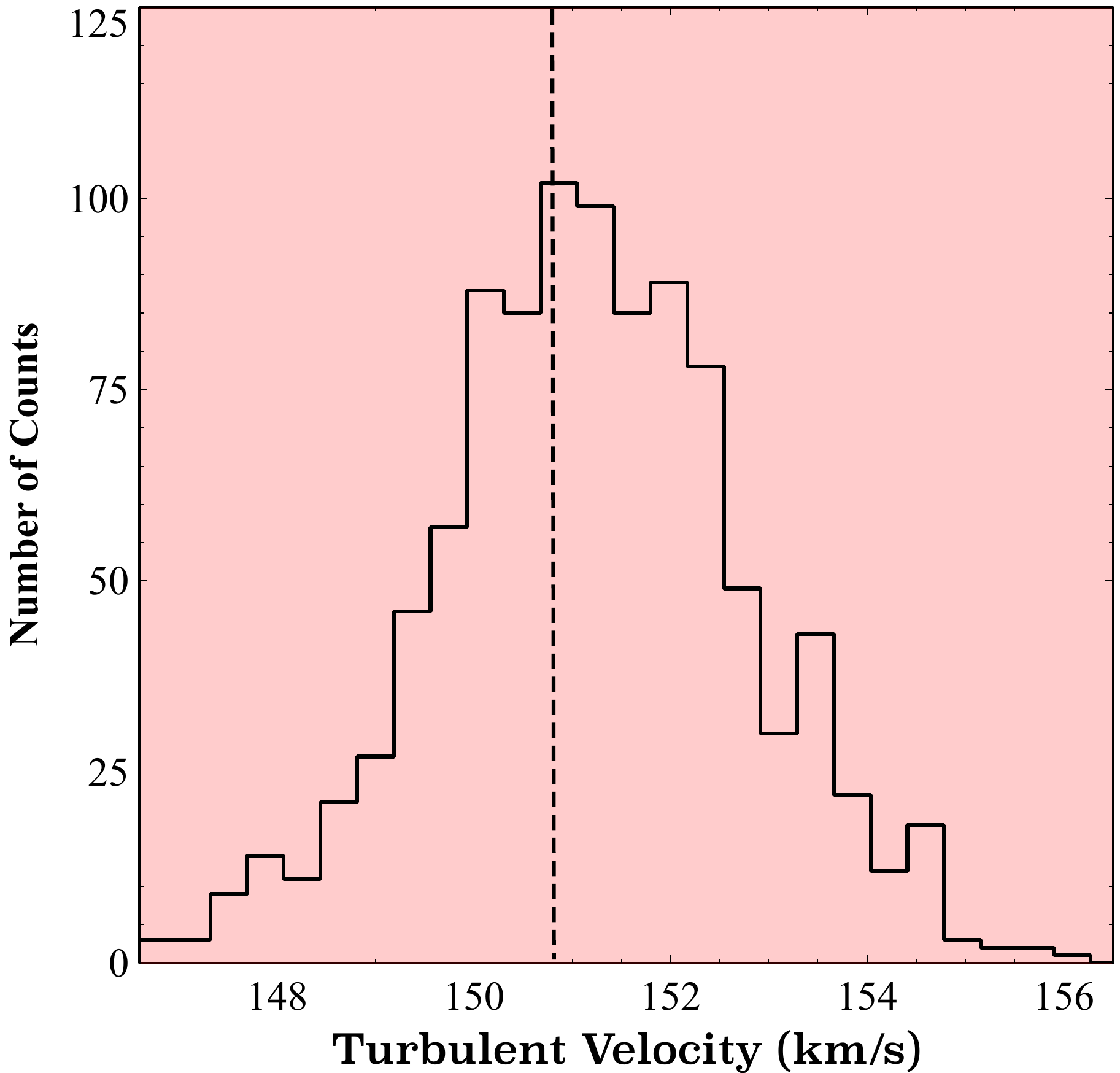}{0.25\textwidth}{(d)}
           }
\gridline{\fig{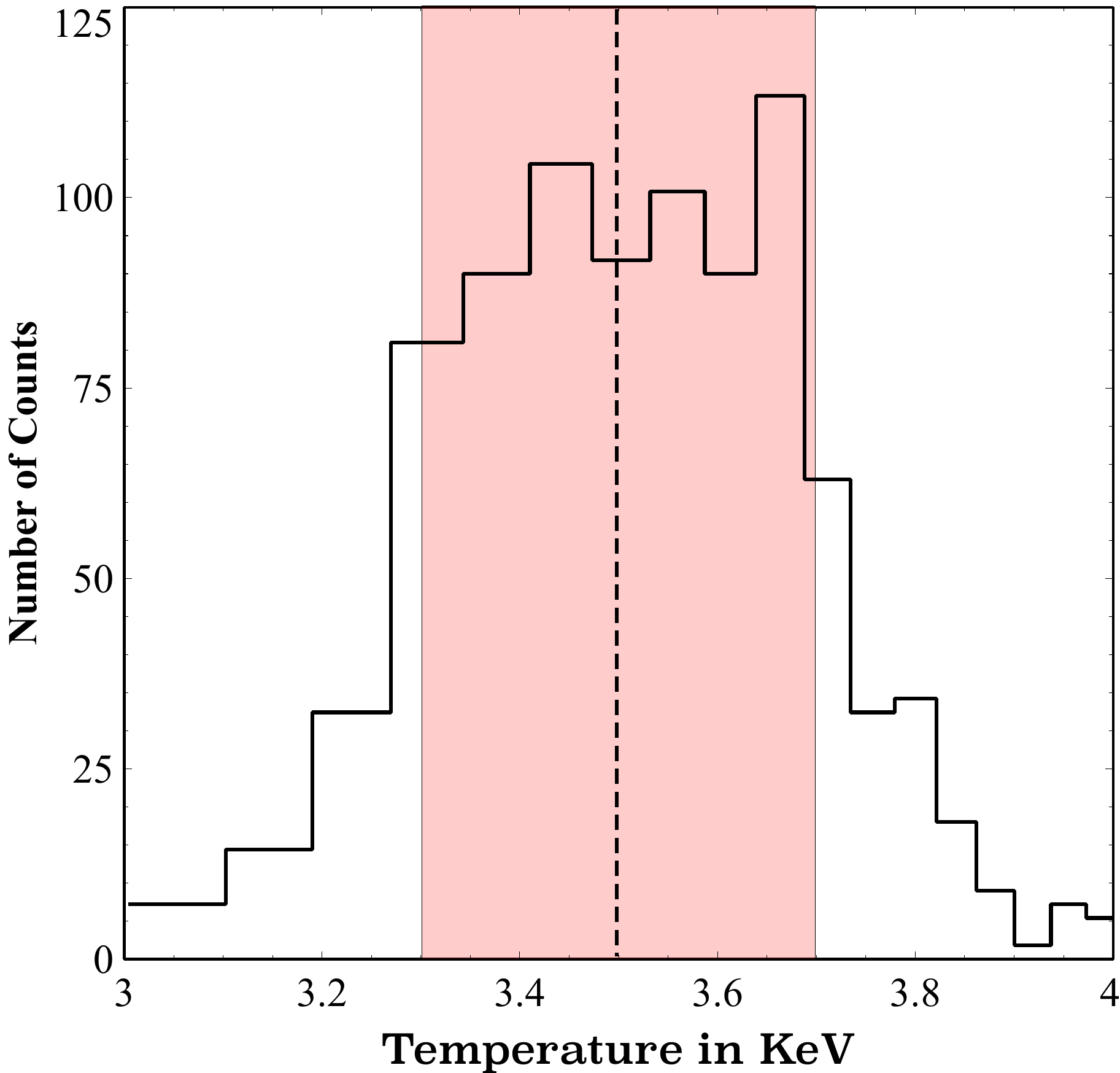}{0.25\textwidth}{(a)}
          \fig{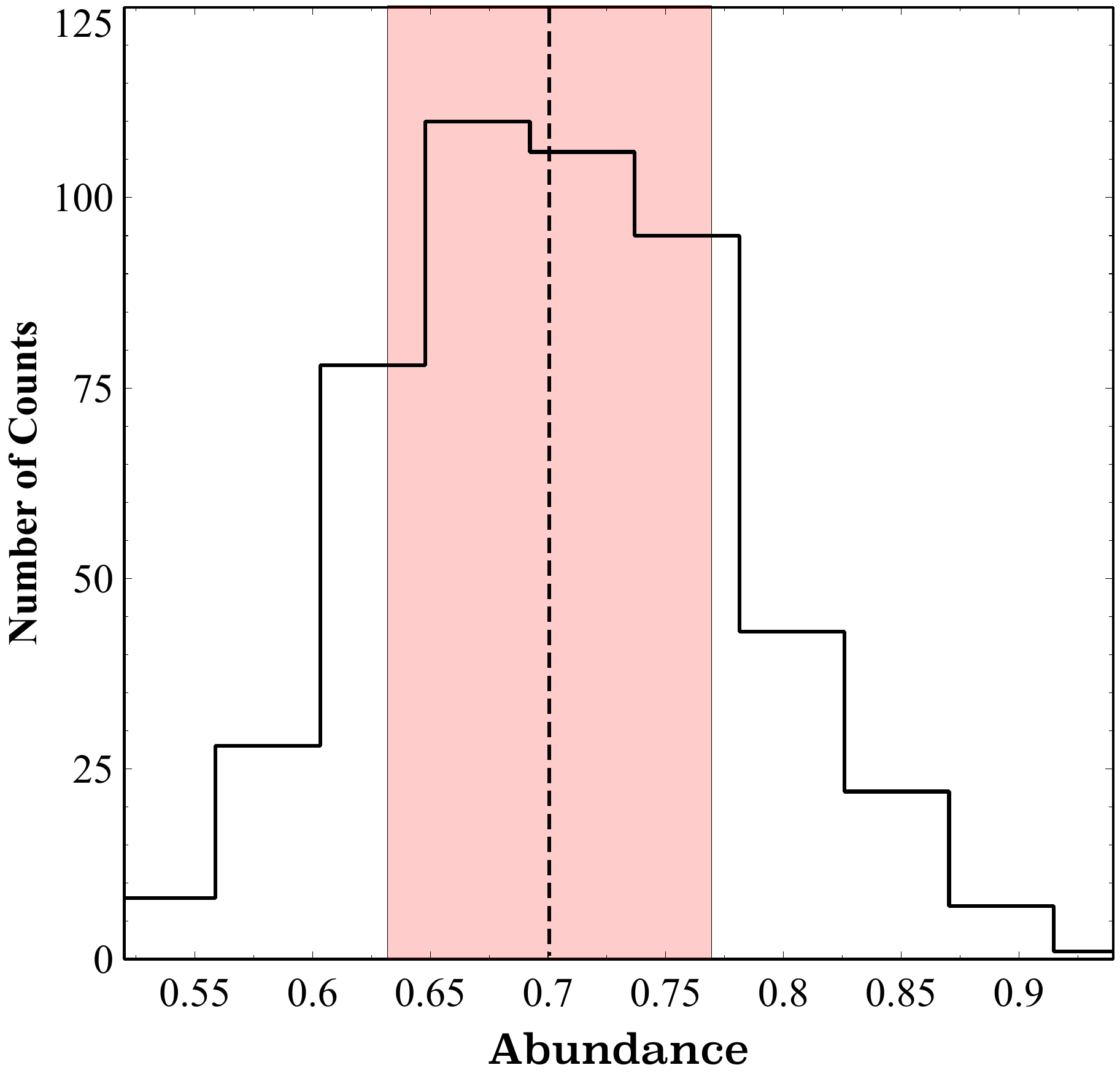}{0.25\textwidth}{(b)}
          \fig{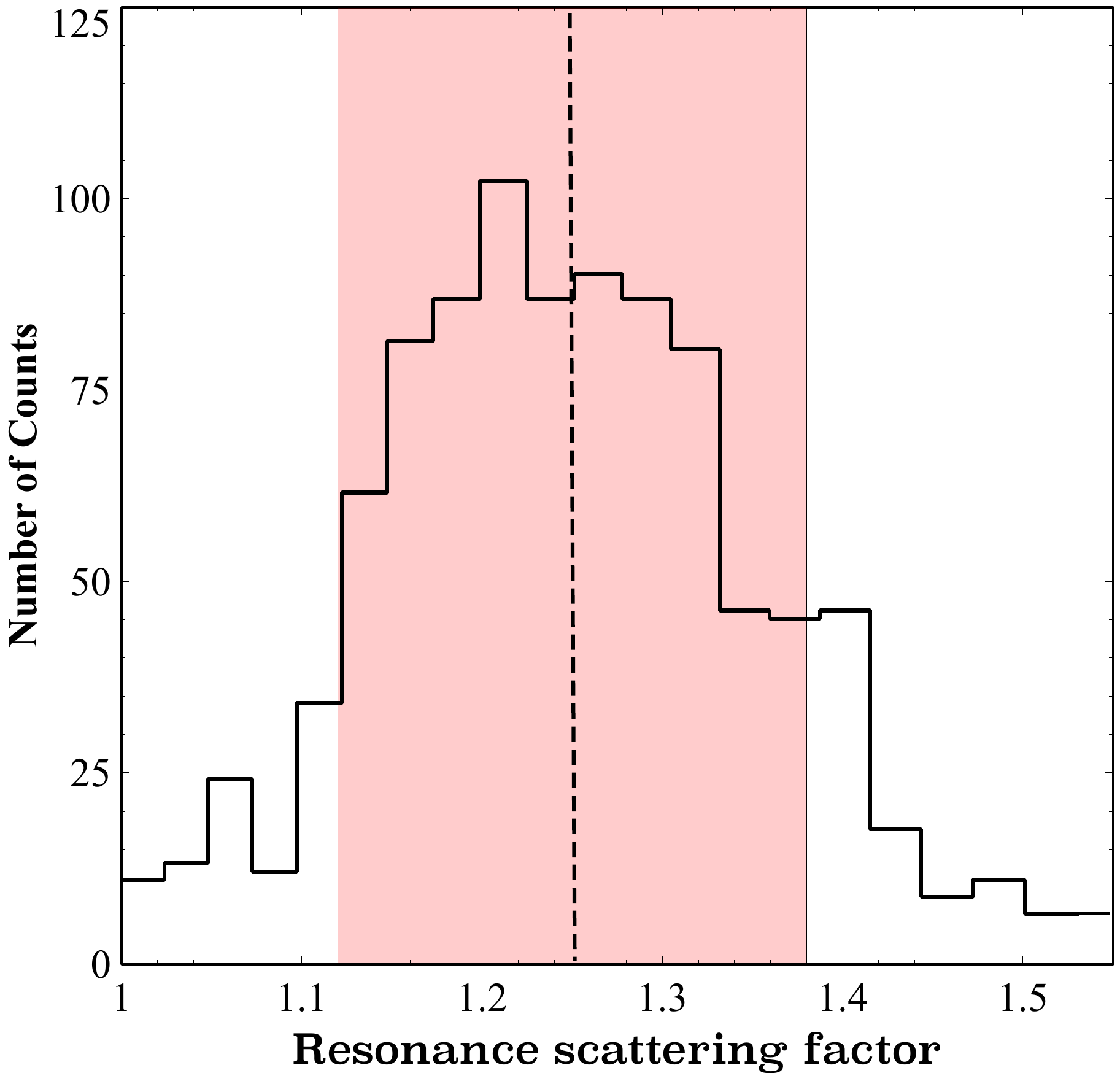}{0.25\textwidth}{(c)}
          \fig{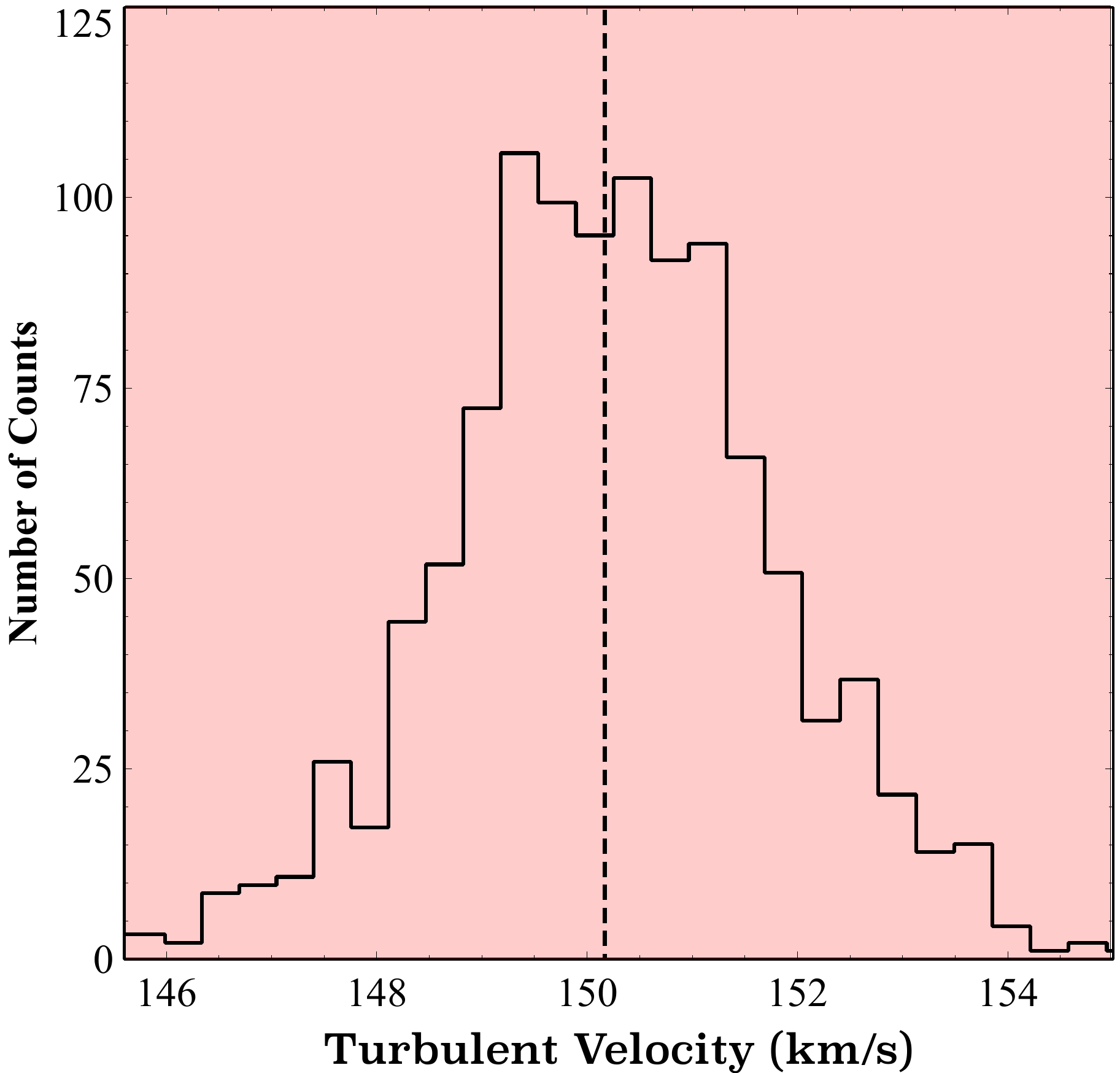}{0.25\textwidth}{(d)}
           }
          
\caption{Top: The distribution of temperature, abundance, resonance scattering factor, and turbulent velocity with varying atomic uncertainties within the energy range 1.8 - 20.0 keV. 
Bottom: The distribution of the same parameters within the energy range 6.0 - 6.8 keV with varying atomic uncertainties fitting the spectrum simulated at the best-fit dominant temperature. The vertical black dashed lines denote the best-fit values (no uncertainty), while red-shaded regions represent their 3$\sigma$ uncertainties.}
\label{fig:20keV}
\end{figure*}

Previous studies have recognized a multi-temperature structure in the Perseus cluster core, with a dominant cooler temperature in the 3-4 keV range when fitted with a two-temperature model \citep{2018PASJ...70...11H}. 
In order to examine whether the uncertainties in Sections \ref{fitting} arise from attempting to fit these with a single temperature as opposed to atomic data issues, we conducted a test on a simulated single-temperature \textit{Hitomi} spectrum.  To initiate this test, we reanalyzed the two-temperature fit for our specified region of interest within the energy range of 6.0-6.8 keV. The resulting best-fit dominant temperature was determined to be 3.50 $\pm$ 0.20 keV. Additionally, the best-fit values for abundance, RS factor, and velocity broadening were found to be 0.70 $\pm$ 0.07, 1.25 $\pm$ 0.13, and  150.23 $\pm$ 8.00 km/s, respectively. Following that, we employed the XSPEC tool \texttt{fakeit} to generate a single-temperature simulated spectrum for a 230 ks observation with \textit{Hitomi}, utilizing the above-derived parameters.  Using the simulated spectrum, we repeated the steps outlined in section \ref{fitting}. The distribution of best-fit temperature, abundance, RS factor, and turbulent velocity is displayed in the bottom panel of Figure \ref{fig:20keV}. The best-fit temperature exhibited a range of 3.00 to 3.99 keV, the best-fit abundance ranged from 0.52 to 0.94, the best-fit RS factor varied between 1.00 to 1.55, and the best-fit turbulent velocity spanned from 145.60 km/s to 155.01 km/s.   
In 30\%, 34\%, and 23\% of the runs, we observe deviations beyond the 3$\sigma$ error bars from the best-fit values for temperature, abundance, and RS factor, respectively.We find 30\%, 34\%, and 23\% of runs the temperature, abundance, and RS factor, respectively, to deviate beyond the 3$\sigma$ error bars from the best-fit values. Therefore, utilizing a simulated one-temperature spectrum leads to a slight decrease in the variability of measured parameters in comparison to fitting the actual spectrum.

\section{Discussion and Conclusion}

The next-generation high-resolution spectroscopic instruments like \textit{XRISM} and \textit{Athena}  minimize systematic errors and provide more precise but not necessarily accurate abundance measurements. In this paper, we present the new capabilities  of \texttt{variableapec}, 
enabling the integration of atomic uncertainties into emissivity calculations.
 The modified emissivities can subsequently be fed into the commonly used \texttt{apec} model, facilitating spectral fitting that accounts for data uncertainties, an essential feature in the data-fitting routines in the era of microcalorimeters.
We also ask the question: Are the best-fit parameters derived from fitting the observed data with the atomic model accurate enough, or do the underlying atomic data within the spectral synthesis models require improved laboratory measurements?

\subsection{Accuracy in temperature}

Accurate temperature measurements provide crucial insights into the structure, gas dynamics, and physical processes occurring in clusters, such as gas entropy and accretion shock in the cluster outskirts \citep{2019SSRv..215....7W, 2021MNRAS.501.3767S, 2022MNRAS.516.3068S}, and merger shock, cold fronts,  and gas sloshing spirals near the cluster center \citep{2023ApJ...944..132S,2022ApJ...935L..23S}.
In galaxies, probing galactic structure, studying the interstellar medium, and understanding stellar populations rely on accurate temperature measurement \citep{2015ApJ...800..102H}. 
Understanding the explosion mechanism, shock physics, and properties of the emitting plasma in SNRs also requires accurate temperature measurement \citep{2012A&ARv..20...49V}. 
 Due to the close correlation between cluster mass and temperature (M $\propto$ T$^{3/2}$), accurate mass measurement in galaxy clusters is closely correlated to accuracy in cluster temperature \citep{1996ApJ...469..494E, 1996A&A...305..756S}. Accurate mass measurement is crucial  for determining the distribution of dark matter, enabling tests of the theory of gravity at large scales, and refining cosmological models \citep{2013ApJ...766...32B}.\\
We ran a thousand simulations varying the atomic data uncertainties within the existing uncertainty in different atomic databases and fitted the \textit{Hitomi} observation of Perseus cluster around the Fe K$\alpha$ complex.
Out of 1000 simulations, including atomic data uncertainties, we found that within the 6.0-6.8 keV energy range $\sim$ 32\% of the best-fit temperatures lie outside the 3$\sigma$ error range of the best-fit temperature determined with zero atomic data uncertainty, implying that improved atomic data is required for accurate temperature measurement. The expansion of the energy range to 1.8-20.0 keV resulted in all temperature values conforming to the 3$\sigma$ error of the best-fit temperature measured with zero atomic data error.

\subsection{Accuracy in abundance}
Accurate abundance measurement is highly significant for constraining the physical environments of the Type Ia (SNIa) and core-collapse (SNcc) supernovae that contributed to the enrichment of the ICM by creating and dispersing metals over the entire volume of clusters \citep{2007A&A...465..345D, 2011MNRAS.418.2744M, 2022MNRAS.516.3068S}. In galaxies, accuracy in the measured chemical abundance is important for studying galactic chemical evolution, characterizing stellar populations, tracing star formation and stellar feedback, and testing models of galactic formation and evolution \citep{2004ApJ...606...92S}. 
Out of 1000 simulations within the 6.0-6.8 keV energy range, in $\sim$ 35\% of the cases, the best-fit abundance lies outside  the 3$\sigma$ 
error regions of the best-fit abundance calculated for zero uncertainty in the atomic data, prompting the need for better laboratory measurements for various collisional rate coefficient and Einstein A coefficients.
The expansion of the energy range to 1.8-20.0 keV led to a slight narrowing of the abundance distribution, with 26\% of abundance values falling outside the 3$\sigma$ error of the best-fit abundance, as determined with zero atomic data uncertainty.

\subsection{Accuracy in RS factor}
Several studies over the years demonstrated that the accurate determination of physical properties in galaxies, clusters, and supernova remnants is heavily reliant on the accuracy of the RS factor, which is also tied to the accurate abundance measurement. Examples are, measurements of velocity fields and gas motion anisotropy \citep{2010SSRv..157..193C}, polarization of X-ray emission lines in elliptical galaxies and galaxy clusters \citep{2010MNRAS.403..129Z}, determination of heavy metal abundances in ICM \citep{2000AdSpR..25..603A}, observability of the hot circumgalactic medium (CGM) in galaxies \citep{2023MNRAS.522.3665N}, and properties of the shell of the Cygnus Loop supernova remnant \citep{2019ApJ...871..234U}. Yet, in the energy range of 6.0-6.8 keV, we found in the 1000 simulations including uncertainties, $\sim$ 25\% of the best-fit RS factors fall outside the 3$\sigma$ uncertainty range of the
best-fit RS factor calculated without uncertainties, requiring further improvement in the collisional rate coefficients of w, y, and z for accurate measurement of the RS factor. Expanding the energy range to 1.8-20.0 keV slightly narrowed the RS factor distribution, with 22\% falling beyond the 3$\sigma$ error of the best-fit RS factor, calculated with zero atomic data uncertainty.

\subsection{Accuracy in turbulence}

The turbulent velocity broadening is inversely related to the  line suppression  in resonance lines via resonant scattering. Like the RS factor, accurate measurement in turbulence is equally important for estimating the velocity fields, density inhomogeneities, and deviation from spherical symmetries \citep{2018PASJ...70...10H}. Turbulent broadening is an integral part of optical depth calculation, useful in estimating cluster mass and concentration, as well as feedback mechanism in Active Galactic Nuclei (AGN)  Supernovae (SNe) \citep{2017ApJ...837..124F}. Best-fit turbulences for all 1000 simulations within the narrow (6.0-6.8 keV) and broad (1.8-20.0 keV) energy ranges with atomic data uncertainties were within the 3$\sigma$ error range of the measured turbulence with no error in the atomic data, implying that the atomic data is accurate enough for precise turbulence measurement.

\subsection{Which atomic data require improvements?}

To determine which atomic data require precise laboratory measurement for acquiring accurate best-fit parameters, we search for correlations of the measured temperature, abundance, RS factor, and turbulence with  Einstein A coefficients and  collisional rate coefficients of individual transitions. We find negative correlations between temperature and z  collisional rate coefficient, abundance and y  collisional rate coefficient, abundance and z  collisional rate coefficient, and a positive correlation between the RS factor and w  collisional rate coefficient, emphasizing the necessity for accurate measurement of these quantities. 
Theoretical determination of collisional rate coefficients assumes that the interaction potential between the colliding systems is perfectly known, which is not the case and constitutes the largest source of error in the rate coefficients with uncertainties that are challenging to assess, prompting the need for lab measurements.
Historically, laboratory measurements have been employed to obtain collisional rate coefficients of highly-ionized ions \citep{1972SSRv...13..565K}.
Theta-pinch is a commonly used device and had been previously used to measure  collisional rate coefficient for all excited states up to 1s$^{2}$5f of the lithium-like silicon ion \citep{1993PhyS...48....9K}. Similar techniques can be utilized for experimentally measuring z,y, and w  collisional rate coefficients.

These models are available at the AtomDB website:http://www.atomdb.org/. 
The necessary information for executing the \texttt{variableapec} is available here: https://github.com/AtomDB/pyatomdb/blob/variableapec/pyatomdb/
pyatomdb/readme\_variableapec.txt

\section*{Acknowledgment}
We acknowledge support by NASA ADAP Grant 80NSSC21K0637.

\bibliography{sample631}{}
\bibliographystyle{aasjournal}



\end{document}